\documentclass[12pt]{article}
\usepackage{a4}
\usepackage{epsfig}

\begin{document}
\title{Spontaneous magnetization of a vacuum in the hot Universe and intergalactic magnetic fields}
\author{
V. Demchik\thanks{e-mail: vadimdi@yahoo.com} {~~\small and~~} V. Skalozub\thanks{e-mail: Skalozubv@daad-alumni.de}\\
{\small Dnipropetrovsk National University, 49010 Dnipropetrovsk, Ukraine}}
\maketitle
\begin{abstract}
We review the spontaneous magnetization of the vacuum of non-Abe\-lian gauge fields at high temperature. The standard model of particles is investigated as a particular example. By using both analytic methods of quantum field theory and gauge field theory on a lattice, we determine the Abelian (chromo)magnetic  fields in the restored phase of the model at high temperatures $T \geq T_{ew}$. The fields are stable and temperature dependent, $B = B(T)$. We investigate the mechanisms of the field stabilization in detail. The screening parameters for electric and magnetic fields -- the Debye, $m_D(B,T),$ and magnetic, $m_{magn.}(B,T)$, masses -- are calculated. It is shown that, in the field presence, the former one is smaller than at zero field. The magnetic mass of the (chromo)magnetic fields is determined to be zero, as for usual $U(1)$ magnetic field. We also show that the vacuum magnetization stops at temperatures below the electroweak phase transition temperature, $T \leq T_{ew}$ , when a scalar condensate creates. These properties make reasonable a possibility that the intergalactic magnetic fields observed recently were spontaneously generated in the hot Universe at the reheating epoch due to vacuum polarization of non-Abelian gauge fields. We present a procedure for estimating the field strengths $B(T)$ at different temperatures. In particular, the value of $B(T_{ew}) \sim 10^{14} G$, at $T_{ew}$ is estimated with taking into consideration the observed intergalactic magnetic field $B_0 \sim 10^{- 15} G$. The magnetic field scale is also estimated. Some model dependent peculiarities of the phenomena studied are briefly discussed.
\end{abstract}

\maketitle

\section{Introduction}
Strong magnetic fields present in all the objects of the observed Universe. As it is believed, these fields   presented  at previous stages of its evolution. They also  are produced in the heavy ion collisions at high energies and influence various processes and characteristics of  plasma, in particular, the deconfinement phase transition temperature. So, searching for  effective mechanisms for generation of different type magnetic fields at high temperature is of great importance for either particle physics or cosmology. These mechanisms could serve as a theoretical background  for investigations of the QCD vacuum at high temperature in magnetic fields \cite{D'Elia:2010nq, Kharzeev:2007jp, Skokov:2009qp, Gatto:2010pt, Aoki:2006we} and the primordial magnetic fields in the early Universe \cite{Giovannini:2012tx, DeSimone:2011ek, Widrow:2011hs, Giovannini:2006sw}. Recent  discovery of extragalactic magnetic fields possessing the field strength of the order $B_0 \sim 10^{-15}G$ at $1 - 10 Mpc$ scale \cite{Ando:2010rb, Neronov:1900zz} is one of the most bright  events of modern cosmology. In \cite{Essey:2010nd} a model-independent 95\% CL interval $1 \cdot 10^{-17} G \leq B \leq 3 \cdot 10^{-14} G $ is determined, and the femtogauss values are actual field strengths in intergalactic space.  Notice that already as the strengths of the intergalactic fields the values $ B \sim  10^{-9} G $ have been  discussed as the most probable ones.  First of all, it makes most reasonable the cosmological origin of primordial magnetic fields. From theoretical point of view, this discovery restricts in an essential way  possible processes supplying the generation of fields in the hot Universe \cite{Neronov:1900zz}. So that  searching for mechanisms of their creation is intensified. Of course, the prime candidates are the primordial fluctuations. But there are more candidates, for a review see \cite{Grasso:2000wj, Giovannini:2003yn, Kandus:2010nw}. A challenge is to produce coherent magnetic fields on different large scales in almost empty intergalactic space.  Different mechanisms operating at different stages of the  evolution and, consequently,  different temperatures are proposed in the literature.  These are trace anomalies, inflation, stochastic electric currents, cosmic strings, paramagnetic resonances, etc., see Refs.~\cite{Giovannini:2000dj, Dolgov:2014gda, Durrer:2013pga, Shtanov:1994ce, Gorbunov:2011zzc}.

In the present paper, we review one of possibilities -- spontaneous magnetization of the vacuum of non-Abelian gauge fields at high temperature. This phenomenon qualitatively (or phenomenologically) is similar to   ferromagnetic media, where  the magnetic domains are formed due to spin interactions of charged particles. In field theory, the role of these particles  plays the quantum fluctuations of  colored  non-Abelian gauge fields having a large magnetic moment (gyromagnetic ratio $\gamma = 2$ of charged gluons and $W$-bosons). Just due to this large value of  magnetic moments   the vacuum magnetization happens. The fields generated spontaneously are long-range, temperature dependent and stable. All these features  make them very attractive objects for various applications. In the frame of the standard model (SM) of particle, the $SU(3)$ color magnetic fields related with strong interactions and $SU(2)_{ew}$  components of usual magnetic field are generated at high temperature. Of course, other type chromomagnetic fields, depending on the gauge groups, could be present  in the early Universe. In what follows, we describe  this phenomenon and, as application,  consider the generation of intergalactic magnetic fields.

The spontaneous magnetization of the vacuum of non-Abelian gauge fields at finite temperature was discovered recently in $SU(2)$ gluodynamics.
It was investigated either by analytic methods in \cite{Starinets:1994vi, Enqvist:1994rm, Skalozub:1996ax, Skalozub:1999bf}
or in lattice simulations in \cite{Demchik:2008zz, Demchik:2007ct}. The basic idea
 rests on the known fact that the spontaneous vacuum magnetization is the consequence of the spectrum of a color charged gluon,
\begin{equation}\label{spectrum}
p^2_{0} = p^2_{||} + (2 n + 1) g
B\qquad(n = - 1, 0, 1,... ),
\end{equation}
in a homogeneous magnetic background, $B$, described by the potential
\begin{equation}\label{potentialB}
 A_\mu^a = B x_2 ~\delta_{\mu 3} \delta^{a 3},
\end{equation}
where $a$ is weak isotopic index, and $p_{||}$ is a momentum component along the field direction. This Abelian type field is the solution to the classical gauge field equations without source terms. Here, a tachyon mode is present in the ground state ($n=-1$). In fact, one observes $p_0^2<0$ resulting from the interaction of the magnetic moment of the spin-1 charged particles with the magnetic field. This phenomenon was firstly discovered by Savvidy \cite{Savvidy:1977as} at zero temperature, $T=0$, and become known as Savvidy vacuum. However, at zero  temperature, this state is not stable. It decays under emission of gluons until the magnetic field $B$ disappears.

This picture changes with increasing of the temperature when a stabilization sets in. The stabilization is due to a vacuum polarization. It depends on two dynamical parameters appearing at $T \not = 0$. These are a magnetic mass of the color charged gluon, $m_{magn.},$ and a $A_0$-condensate, which is proportional to the Polyakov loop \cite{Ebert:1996tj}. This field configuration is stable, its energy is below the perturbation one and the minimum is reached for the field strengths  of order $g B \sim g^4 T^2/\log T$. This phenomenon is common for different $SU(N)$ gauge groups which can be used to extend the standard $(SU(2) \times U(1))_{ew} \times SU(3)_c$ model of elementary particles.

An important property of such temperature dependent magnetic fields is the vanishing of their magnetic mass, $m_{magn.} = 0$. It was found both in one-loop analytic calculations \cite{Bordag:2006pr} and lattice simulations \cite{Antropov:2010tj}. The mass parameter describes the inverse spatial scales of the transverse field components, similarly to the Debye mass $m_D$ related to the inverse space scale for the electric (Coulomb) component. The absence of the screening mass means that the spontaneously generated Abelian chromomagnetic fields are long-range at high temperature, as it is common for the $U(1)$ magnetic field. Hence, it is reasonable to believe that, in the hot Universe, at each stage of its evolution spontaneously created, strong, long-range magnetic fields of different types have been present. The fields influenced various processes and phase transitions.

The aim of the present paper is to describe the main points of the vacuum magnetization phenomenon, investigate the properties of the created fields and discuss some possible physical consequences of it. In fact, this phenomenon is a non-perturbative one. So, the main results have been obtained both in continuum field theory  and in the lattice simulations.  The results  are in agreement with each other. Every method of  calculations mentioned has specific features which will be described below. We consider the magnetization in the SM as well as the minimal supersymmetric standard model (MSSM) at high temperatures and compare the results. We describe the mechanisms resulting in stabilization of the magnetized vacuum. As an application, we consider the origin of intergalactic magnetic fields at the reheating stage  of the universe evolution that will be investigated in more details. It worth mentioning  that one of  difficult problems of magnetogenesis is to relate the field strengths generated in the early Universe with the present day fields, that depends on numerous factors and is model dependent. These points will be  discussed also.

In the next section, we start with analytic calculations and describe the consistent effective potential (EP) $V(T, B, \phi_c) $ of magnetic and scalar fields used in studying of the spontaneous vacuum magnetization in the SM. The main observation here is that the magnetization does not happen when a sufficiently large scalar field condensate $\phi_c \not = 0$ is present. It means, in particular, that after electroweak phase transition (EWPT) the spontaneous vacuum magnetization does not happen. This concerns the usual magnetic field related with $SU(2)_{ew}$ gauge group. Chromomagnetic field does not couple to $\phi_c$ and therefore may exist till the deconfinement phase transition  temperature $T_d$.  In what follows, we consider the magnetization in different models of particles for the case of $\phi_c = 0$, only. Sect.~3 is devoted to investigation of the vacuum magnetization for the SM. The temperature masses in the field presence is discussed in Sect.~4. Sects.~5, 6 are dealing with  calculation of the Debye and magnetic masses for neutral and charged gluons at high temperature in the external magnetic field presence. For these purposes the one-loop gluon polarization tensors are calculated and investigated. Then in Sects.~7 and 8 we turn to non-perturbative methods  and investigate the vacuum magnetization and the magnetic mass of neutral gluons at finite temperature in the field presence by using the Monte Carlo (MC) methods on the lattice. We obtain the results coinciding with that of the analytic calculations. In Sect.~9 we develop a procedure for  relating  magnetic fields generated in the hot Universe with the present day magnetic fields and estimate the field strength at the EWPT temperature $T_{ew}$. In Sect.~10 the scale of magnetic fields at different temperatures is estimated. Discussion of the results obtained and prospects is given in the final section.

\section{Effective potential at finite temperature}
As we noted above, the spontaneous vacuum magnetization and the absence of the magnetic mass for the Abelian magnetic fields are non-perturbative effects to be determined, in particular, in lattice simulations \cite{Demchik:2008zz, Antropov:2010tj}. The main conclusions of these investigations are that the stable magnetized vacuum does exist at high temperature and that the magnetic mass of the created field is zero. Concerning the actual value of the field strength, it is close to the one calculated within the consistent effective potential which takes into account the one-loop plus daisy diagrams. So, in analytic calculations we restrict ourselves to this approximation.

\begin{figure}
\begin{center}
\includegraphics[bb=0 0 540 123,width=0.5\textwidth]{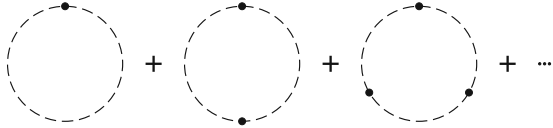}
\end{center}
\caption{The neutral gauge fields and scalar field ring diagrams giving contributions to the effective potential.}\label{figringneutral}
\end{figure}

\begin{figure}
\begin{center}
\includegraphics[bb=0 0 540 130,width=0.5\textwidth]{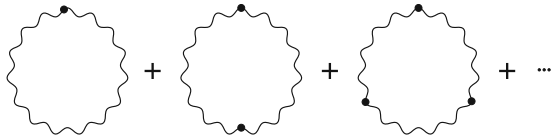}
\end{center}
\caption{The charged gauge field ring diagrams giving contributions to the effective potential.}\label{figringcharged}
\end{figure}

The complete EP for the SM is given in  \cite{Skalozub:1999pw}. Its general structure reads
\begin{equation}\label{EP}
 V(T, B, \phi_c) = \frac{B^2}{2} + V^{tree}(\phi_c) + V^{(1)}(T, B, \phi_c) + V^{ring}(T, B, \phi_c),
\end{equation}
where $\phi_c $ is a scalar field condensate. The first two terms describe the tree level contributions of classical magnetic and scalar fields, $V^{(1)}(T, B, \phi_c)$ is the one-loop contributions of all the fields and the last term presents the contributions of ring (or daisy) diagrams giving the main long-range correlation corrections.
To find $V^{ring}(T, B, \phi_c)$ the one-loop polarization operators of charged, $\Pi(T, B)$ and neutral, $\Pi^0(T, B)$ gluons in the external field and at finite temperature have to be calculated in the limit of zero momenta. Then $V^{ring}(T, B)$ is given by the series depicted in figures \ref{figringneutral}, \ref{figringcharged}. Here, the dashed lines
describe the neutral gluons and the wavy lines represent the charged ones, blobs stand for the one-loop polarization operators. The diagrams with one blob show the two-loop terms of the EP, with two blobs - three-loop ones, etc.

The one-loop terms have the orders $\alpha_{em}= e^2/(4 \pi)$ and $ \lambda$ in a fine structure constant and a scalar self-interaction coupling. The order of the rings is $\alpha_{em}^{3/2}, \lambda^{3/2}$, correspondingly. So that these diagrams are dominant compared to the two-loop contributions having the order $\alpha_{em}^{2}, \lambda^{2}$. As it is well known, the EP (\ref{EP}) is a consistent approximation because it is real at sufficiently high temperature (see for details \cite{Skalozub:1999pw} and references therein). So, it will be used in the following.

In the present study we are interested in two limits of it:
\begin{enumerate}
 \item weak magnetic field and large scalar field condensate, $e B < \phi^2_c, T \not = 0$,
 \item the case of the restored symmetry, $\phi_c = 0, B \not = 0, T \not = 0$.
\end{enumerate}
For the former case we show the absence of spontaneous vacuum magnetization at finite temperature. For the latter one we estimate the field strength at high temperature.

To demonstrate the first property we consider the one-loop contribution of $W$-bosons calculated in a standard way as the sum over the spectrum (\ref{spectrum}) (see Ref.~\cite{Skalozub:1999pw} for details):
\begin{eqnarray}\label{L2t}
 V^{(1)}_w(T,h,\phi) &=& \frac{h}{\pi^2 \beta^2} \sum\limits_{n = 1}^{\infty} \Bigl[ \frac{(\phi^2 - h)^{1/2}\beta}{n} K_1(n \beta (\phi^2 - h)^{1/2}) \\ \nonumber
 &-& \frac{(\phi^2 + h)^{1/2}\beta}{n} K_1(n \beta (\phi^2 + h)^{1/2})\Bigr].
\end{eqnarray}
$n$ labels discrete energy values and $K_1(z)$ is the MacDonald function. Here and in what follows we use the dimensionless variables: $h = e B/M^2_w$, $\phi = \phi/\phi_0$ ($\phi_0$ is the value of scalar field condensate at zero temperature), and $\beta = M_w/T$, $M_w$ is the $W$-boson mass. In fact, other terms of the EP have to be added. However, to elucidate the role of the scalar field condensate this term is sufficient.

Now, let us show that the spontaneous vacuum magnetization does not happen at finite temperature and for non small values of the scalar field condensate $\phi \not = 0$. To do that we notice that the magnetization is produced by the gauge field contribution, given in Eq.~(\ref{L2t}). So, we consider the limit of $\frac{e B}{T^2} \ll 1$ and $\phi^2 > h$. For this case we use the asymptotic expansion of $K_1(z)$,
\begin{equation}\label{K1asympt}
 K_1(z) \sim \sqrt{\frac{\pi}{2 z}} e^{- z} \left( 1 + \frac{3}{8 z} - \frac{15}{128 z^2} + \cdots \right),
\end{equation}
where $ z = n \beta (\phi^2 \pm h)^{1/2}$. Now, we investigate the limit of $\beta \to \infty, \frac{1}{\beta \phi} \ll 1$. For this case the leading contribution is given by the first term of the temperature sum in Eq.~(\ref{L2t}). We can also substitute $(\phi^2 \pm h)^{1/2} = \phi ( 1 \pm \frac{ h}{2 \phi^2})$. In this approximation, the sum of the tree level energy and (\ref{L2t}) reads
\begin{equation}\label{L2tasympt}
 \it{V} = \frac{h^2}{2} - \frac{h^2}{\pi^{3/2}} \frac{1}{\beta^{1/2} \phi^{1/2}} \left( 1 - \frac{1}{2 \beta \phi} \right) e^{- \beta \phi}.
\end{equation}
The second term is exponentially small and the stationary equation $\frac{\partial{V}}{\partial{h}} = 0$ has the trivial solution $h = 0$. This estimate can be verified easily in numeric calculation for the total EP. Hence, we conclude that after symmetry breaking the spontaneous vacuum magnetization does not take place, as at zero temperature \cite{Ghoroku:1982pb}.

The main goal of our investigation is the restored phase of the SM. To begin, we adduce the high temperature
 contribution of the complete effective potential relevant for this case using the results in ~\cite{Skalozub:1999pw}. First
  we write down the one-loop $W$-boson contribution as the sum of the pure Yang-Mills weak-isospin part $(\tilde{B}\equiv B^{(3)}$),
\begin{eqnarray}\label{VW}
 V^{(1)}_w(\tilde{B}, T) = \frac{\tilde{B}^2}{2} + \frac{11}{24} \frac{g^2}{\pi^2} \tilde{B}^2 \log \frac{T}{\mu} - \frac{1}{3} \frac{( g \tilde{B})^{3/2} T }{\pi}
 - i \frac{( g \tilde{B})^{3/2} T }{2 \pi} + O (g^2 \tilde{B}^2),
\end{eqnarray}
where $g$ is weak isospin gauge coupling constant, $\mu$ is a temperature normalization point, and the charged
 scalars \cite{Skalozub:1996ax}
\begin{equation}\label{Vscalar}
 V^{(1)}_{sc}(\tilde{B}, T) = - \frac{1}{48} \frac{g^2}{\pi^2} \tilde{B}^2 \log \frac{T}{\mu} + \frac{1}{12} \frac{( g \tilde{B})^{3/2} T }{\pi} + O (g^2 \tilde{B}^2),
\end{equation}
describing the contribution of longitudinal vector components. Remind that electric charge is expressed through $g$ and the Weinberg angle as $e = g \sin \theta_W$. The first term in Eq.~(\ref{VW}) is the tree-level energy of the field. This representation is convenient for the case of extended models including other gauge and scalar fields. Dependently on a
specific case, one can take into consideration the parts (\ref{VW}), (\ref{Vscalar}), correspondingly. In the SM, the contribution of Eq.~(\ref{Vscalar}) has to be taken with the factor 2, due to two charged scalar fields entering the scalar doublet of the model. In the case of the Two-Higgs-Doublet SM, this factor must be 4, etc. The imaginary part is generated because of the unstable mode in the spectrum (\ref{spectrum}). It is canceled by the term appearing in the contribution of the ring (daisy) diagrams for
the unstable mode \cite{Skalozub:1999bf}
\begin{equation}\label{ringdaisy}
 V_{unstable} = \frac{g \tilde{B} T}{2 \pi} \left[\Pi(\tilde{B}, T, n = - 1) - g \tilde{B} \right]^{1/2} + i \frac{(g \tilde{B})^{3/2} T}{2 \pi}.
\end{equation}
Here, $\Pi(\tilde{B}, T, n = - 1)$ is the mean value for the charged gluon polarization tensor taken in the ground state $n = - 1$ of the spectrum (\ref{spectrum}). A more presize definition is given below. If this value is sufficiently large, spectrum stabilization due to radiation correction is realized. The cancelation of imaginary terms in the sum of Eqs.~(\ref{VW}) and (\ref{ringdaisy})  and, hence,  the vacuum stabilization is similar to that in scalar field theory at finite temperature \cite{Dolan:1973qd, Linde:2005ht}. It means that the one-loop plus ring diagram contributions  give a consistent effective potential at high temperature. Moreover, the ring diagrams have the order $g^3$ (or $\lambda^{3/2}$ is scalar theory). Therefore for small couplings they are more important as  two-loop contributions having the orders $g^4$ and  $\lambda^{2}$, correspondingly.

In the gauge field case, the possibility of stabilization formally follows from the temperature and field dependence of the polarization tensor in the high temperature limit $T \to \infty $ \cite{Bordag:2008wp}: $\Pi(\tilde{B}, T, n = - 1)= c~ g^2 T \sqrt{g \tilde{B}} $, where $c > 0$ is a constant which must be calculated explicitly. At high temperature the first term can be larger than $g \tilde{B}$. Thus, the accounting for of rings leads to the vacuum stabilization. It is obvious in the high temperature limit. But it  also is the case for non-asymptotic temperatures. This is discussed in more details in the next section. It worth to mention that for gauge fields there are two types of rings. First is related with longitudinal electric components of $\Pi_{00}$. Second takes into consideration the properties of the transversal long range modes. Just  the contribution of the latter modes is given in Eq.~(\ref{ringdaisy}).

The high temperature limit of the fermion contribution looks as follows \cite{Skalozub:1999pw},
\begin{equation}\label{fermionEP}
 V_{fermion} = - \frac{\alpha}{\pi} \sum\limits_{f} \frac{1}{6} q^2_f \tilde{B}^2 \log\frac{ T}{\mu},
\end{equation}
where the sum is extended to all the leptons and quarks, and $q_{f}$ is the fermion electric charge in positron units. Hence it follows that in the restored phase for $\mu \gg T_{ew}$ all the fermions give the same contributions.

Now, let us present the EP for the ring diagrams describing the long-range correlation corrections of longitudinal gauge field
components and scalars
at finite temperature \cite{Carrington:1991hz, Demchik:2002ks},
\begin{eqnarray}\label{Vring}
 V_{ring} &=& \frac{1}{24 \beta^2} \Pi_{00}(0) - \frac{1}{12 \pi \beta} {~\rm Tr~} [\Pi_{00}(0)]^{3/2} \\\nonumber
 &&+ \frac{(\Pi_{00}(0))^2}{32 \pi^2} \left[\log\left(\frac{4 \pi}{\beta (\Pi_{00}(0))^{1/2}}\right) + \frac{3}{4} - \gamma \right],
\end{eqnarray}
where the trace means summation over all the contributing states, $\Pi_{00} = \Pi_{\phi}(k = 0, T, B)$ for the Higgs particle; $m_D^2 = \Pi_{00} = \Pi_{00}(k = 0, T, B)$ are the zero-zero components of the polarization functions of gauge fields in the magnetic field taken at zero momenta, called the Debye mass squared, $\gamma$ is Euler gamma. These terms are of the order $\sim g^3 (\lambda^{3/2})$ in the coupling constants. The detailed calculation of these functions is given in ~\cite{Skalozub:1999pw}. We
 give the results for completeness:
\begin{eqnarray}\label{Piscalar}
 \Pi_{\phi}(0) &=& \frac{1}{24 \beta^2} \left( 6 \lambda + \frac{6 e^2}{\sin^2 (2\theta_W)} + \frac{3 e^3}{\sin^2 \theta_W} \right)\\\nonumber
 &&+ \frac{2 \alpha}{\pi} \sum\limits_f \left[ \frac{\pi^2 K_f}{3 \beta^2} - |q_f B| K_f \right]
 + \frac{(e B)^{1/2}}{8 \pi \sin^2 \theta_W \beta} e^2 \left(3 \sqrt{2} \zeta(-\frac{1}{2}, \frac{1}{2})\right).
\end{eqnarray}
Here $K_f = \frac{m_f^2}{M_w^2} = \frac{G^2_{Yukawa}}{g^2}$ and $\lambda$ is the scalar field coupling. The terms \mbox{$\sim \!T^2$} give standard contributions to temperature mass squared coming from the boson and fermion sectors. The $B$-dependent terms are negative (the value of $
3 \sqrt{2} \zeta(-\frac{1}{2}, \frac{1}{2}) = - 0.39$). They decrease the value of the screening mass
at high temperature. The Debye masses squared for the photons, $Z$-bosons and neutral current contributions are, correspondingly,
\begin{eqnarray}\label{mAZ}
 m^2_{D, \gamma} &=& g^2 \sin^2\theta_W \frac{1}{3 \beta^2} + O(e B \beta^2), \\ \nonumber
 m^2_{D, Z} &=& g^2 \left(\tan^2\theta_W + \frac{1}{4 \cos^2\theta_W} \right) \frac{1}{3 \beta^2} + O(e B \beta^2), \\ \nonumber
 m^2_{D,neutral} &=& \frac{g^2}{ 8 \cos^2\theta_W \beta^2} \left(1 + 4 \sin^4 \theta_W\right) + O(e B \beta^2).
\end{eqnarray}
As we see, the dependence on $B$ appears in the order $O(T^{-2})$.

The $W$-boson contribution to the Debye mass of the photons is
\begin{equation}\label{mW}
 m^2_{D, W} = 3 g^2 \sin^2\theta_W \left(\frac{1}{3 \beta^2} - \frac{(g \sin \theta_W B)^{1/2}}{2 \pi \beta}\right).
\end{equation}
Interesting feature of this expression is the negative sign of the next-to-leading terms dependent on the field strength.

Finally, we give the contribution of the high temperature part in Eq.~(\ref{ringdaisy}) $\Pi(\tilde{B}, T, n = - 1)$,
\begin{equation}\label{Piunstable}
 \Pi(\tilde{B}, T, n = - 1) =  \alpha \left[\frac{3.26}{\sqrt{4 \pi}} \frac{(g \sin \theta_W B)^{1/2}}{\beta} + i \frac{(g \sin \theta_W B)^{1/2}}{\beta} \right].
\end{equation}
This expression was calculated from the space components of the one-loop $W$-boson polarization tensor in the external field at high temperature (see Eqs.~(\ref{mmass}), (\ref{rmass}) below). In what follows, we use an imaginary time formalism and the Schwinger-Dyson equation in the operator form $D^{-1} = \Delta^{-1} - \Pi$, where $\Delta$ is tree-level propagator. In this notation the function $\Pi(\tilde{B}, T, n = - 1) = - <n = - 1| \Pi  | n = - 1 >$. In contrast to Debye's mass, it is proportional to the   product  $T \sqrt{B}$ and called "magnetic mass".  It is one of the main objects for the problem of interest. We consider  calculation of both neutral and charged modes of it in Sects.~5, 6. The expression (\ref{Piunstable}) contains the imaginary part which comes from the unstable mode in the spectrum (\ref{spectrum}). Its value is of the order of the usual dumping constants in plasma at high temperature. Of course, the origin of these imaginary terms is different.   It will be  ignored in actual calculations in what follows. This is because we are interested mainly in the high temperature case where the imaginary part remaining after cancelation of the  imaginary terms in Eqs.~(\ref{VW}) and (\ref{ringdaisy}) is of the next-to-leading order in the consistent effective potential.

In fact, this part must be calculated in a more consistent scheme which starts with a regularized stable spectrum. The stabilization can be realized not only by the radiation corrections but also some other mechanisms. We observed a stable vacuum state in the lattice simulations \cite{Demchik:2008zz}. Therefore, we believe that this problem has a positive solution. Other important observation obtained is the absence of the vacuum magnetization at not small values of the scalar field condensate, $\phi_c \not = 0$. In particular, this means that just after the electroweak phase transition this phenomenon does not hold.

\section{Vacuum magnetization at high temperature}
The EP (\ref{L2t}) (and other terms in ~\cite{Skalozub:1999pw}) is expressed through the well known special functions. Therefore, it can easily be investigated numerically for any range of parameters entering. As usually, it is convenient to introduce the dimensionless variables: the field $\Phi = (g \tilde{B})^{1/2}/T$ and the EP $v(\Phi,g) = V(\tilde{B},T)/T^4$. The vacuum magnetization at high temperatures, $ T>> (g \tilde{B})^{1/2},\Phi \rightarrow 0$, can be investigated within the following limiting form of the total EP,
\begin{equation}\label{Vlim}
 v^{total}(\Phi,g)_{\mid \Phi\rightarrow 0} = \frac{\Phi^4}{2 g^2} + \left(\frac{5}{3} - \sum_f \frac{1}{6} q^2_f\right) \frac{\Phi^4}{4 \pi^2}
 \log\left(\frac{T}{\mu}\right) - \frac{1}{3} \frac{\Phi^{3}}{\pi}
 + O(g^{2}).
\end{equation}
where $\sim g^{2}$ terms were omitted. The logarithmic term is signaling asymptotic freedom in the field at high temperatures. In this approximation, the role of the ring diagrams is reduced to the cancelation of the imaginary part in Eq.~(\ref{VW}), only. The first term in Eq.~(\ref{ringdaisy}) has the order $\sim g^{9/4}$
and can be neglected for small $g$. This follows from the high temperature estimate
for $\Pi(T, n = - 1)|_{T \to \infty} \sim g^2 T \sqrt{g \tilde{B}}$ (see Eq.~(\ref{mmass}) below).
 Thus, in given approximation the EP is real. From Eq.~(\ref{Vlim}) we obtain for the minimum position
\begin{equation}\label{PhiT}
 (g\tilde{B})^{1/2}_{c}(T) = \frac{g^2 T}{2 \pi}\frac{1}{1 + (\frac{5}{3} - \sum_f \frac{1}{6} q^2_f) \frac{g^2}{4 \pi^2} \log{\frac{T}{\mu}}}.
\end{equation}
Hence, we come to the conclusion that the ferromagnetic vacuum state does exist at high temperatures. The field strength (\ref{PhiT}) is proportional to the coupling constant $g^2$. Such type dependence is opposite to the case of the usual spontaneous breaking of symmetry where the value of the scalar condensate depends non-analytically on the coupling: $\phi_0 \sim 1/\sqrt{\lambda}$.

Let us discuss in details the stability of the condensed field. First we note, for the specific value of the field strength (\ref{PhiT}) the order of the terms in Eq.~(\ref{Piunstable}) is $\sim g^4$, that approves the consistency of the approximation used. Further, the second derivative of the EP is positive for $g \tilde{B}_c$ that means we have a minimum. The field is not changing in the direction $a = 3$ of the isotopic space. To check is this the case or not for the perpendicular directions $ = 1, 2$ or $a = a^{\pm}$ which is responsible for excitation of charged fields $W^{\pm}$, one has to calculate the effective mass squared $M^2(\tilde{B}_c, T)$. Here, we consider the one-loop case. Substituting the value $g \tilde{B}_c^{(1)} $ Eq.~(\ref{PhiT}) in the one-loop polarization function Eq.~(\ref{Piunstable}), we find that the effective mass squared, $M^2(\tilde{B}_c^{(1)}, T) =  [{\rm Re} \Pi^{(1)}(\tilde{B}_c^{(1)}, n = - 1) - g \tilde{B}_c^{(1)}] \ge 0$, is positive for any  ${\rm Re} \Pi$ at sufficiently high temperature. This is because the second term is logarithmic suppressed ($\sim 1/\log(T/\mu))$ compared to the first one. Thus, the vacuum stabilization is expected in this consistent calculation.

The above analytic investigation unambiguously determined the possibility of the vacuum magnetization at high temperature, although a number of questions has to be studied in order to derive a final picture. In fact, the radiation corrections are not the only candidate for stabilization of the vacuum. Other factor is a so-called $A_0$-condensate which is proportional to Polyakov's loop. This parameter also acts as a stabilizing mass term at high temperature \cite{Starinets:1994vi, Meisinger:2002ji, Demchik:2007ct}. Its role will be discussed in what follows. Let us note once again that we are dealing with infrared properties of gauge fields at high temperature. So,  non-perturbative methods of calculations have to be used.  Below, we investigate the magnetization as well as the  field characteristics by means of analytic methods and simulations on a lattice. The coincidence  of the results obtained by using these methods is very important.  The former are needed   in determining of general analytic properties for phenomenons investigated. The latter admit obtaining of  the non-perturbative results.

\section{Temperature masses in the field presence}
As it is well known, at finite temperature particles acquire the temperature dependent masses which modify  interactions in plasma \cite{Kalashnikov:1982sc, Landsman:1986uw}. At zero fields, this mass is of the order $\sim g T$ ($m^2_D = - \Pi_{44}(T, k_4 = 0, \vec{k} \to 0)$) for quarks and longitudinal gauge field excitations. For transversal gluons, this mass is of the order $\sim g^2 T$ ($m^2_{magn.} =  - \Pi_{ij}(T, k_4 = 0, \vec{k} \to 0)$). So that both chromoelectric and chromomagnetic fields become screened in hot QCD. These basic characteristic parameters of quark-gluon plasma (QGP) are very important. They determine the screening properties and the plasma excitations for different type gauge fields. These parameters have been calculated in analytic quantum field theory at finite temperature and simulations on the lattice. Note that in scalar and spinor QED the longitudinal (electric, Debye) mass has the order $\sim e T$ and the magnetic mass of photons equals to zero \cite{Kalashnikov:1982sc}. Therefore the latter  field remains  not screened long range one as in the vacuum.

At non-zero external magnetic fields, these characteristics are not well investigated at present. This some time leads to confusions in the literature where instead of the magnetic mass the Debye mass is substituted \cite{Enqvist:1994rm, Meisinger:2002ji}. That influences the results obtained.  As it was discovered in analytic calculations \cite{Bordag:2006pr} and confirmed in simulations on the lattice \cite{Antropov:2010tj} the magnetic mass of neutral gluons calculated in the external chromomagnetic Abelian field $B = const$ equals to zero. Therefore this field behaves as usual magnetic field in QED. It is long range one, in contrast to the zero field case. The Debye mass of these fields is field-dependent, and its value decreases with growth of the field strength \cite{Bordag:2006pr}. Hence, the color Coulomb field becomes less screened  compared to the zero field case.

The Debye mass of charged gluons (calculated in $SU(2)$ gluodynamics) is also field-depended and coinciding with the neutral gluon value \cite{Bordag:2008zza}. As concerns the magnetic mass for charged gluons, it remains not well calculated. Only rough estimates have been obtained already \cite{Skalozub:1999bf}. But this parameter is of paramount importance for the plasma.

From dimensional analysis it follows that the PT at high temperature has the form $\Pi^{ch.}_{i j}(B, T, p_3 \to 0) \sim c \sqrt{g B} T$, where $c$ is a constant which has to be computed. So that if $c \le 0$ the spectrum stabilization at high temperature ($T \to \infty$) holds. If $c \ge 0$, the spectrum and, therefore, external field remains unstable as at zero temperature.

In fact, we have to distinguish two situations: 1) the spectrum  can be positive for any field strengths at high temperature for $c \le 0$; 2) the spectrum in the field which is spontaneously generated at high temperature ($B = B(T)$). In the later case the value of the gluon effective mass
\begin{equation}\nonumber
m_{eff.}^2 = - g B - \langle n = -1| \Pi^{ch.}_{i j}(p_4 = 0, B, T, p_3 \to 0)| n = -1\rangle
\end{equation}
can be negative dependently on the value of $B(T)$. The fist case is of interest, for example, for the QGP in $pp$-collisions, when one can expect strong external chromomagnetic fields at high temperature produced by quark currents. The second case is of interest for cosmology at $T \ge T_d$.

\section{Polarization tensor of neutral gauge fields at high temperature}
In this section, we investigate the properties of  neutral gauge fields in the restored phase of the SM by analytic methods. As before, in actual calculations we concentrate on the electroweak sector $SU(2) \times U(1)_Y$ of gauge fields. The color $SU(3)_c$ gluons possess  similar properties which, for short, we note, only.

Before  proceeding in actual calculation, we say a few words about  analytic calculations at finite temperature in the fields with the unstable mode in the spectrum (\ref{spectrum}). At zero temperature,  first of all one has to determine  new vacuum state which follows from the instability evolution due to  self-interactions of charged gauge fields. Then, quantum excitations at the stable vacuum should be investigated. As also is known, in this case the color external field is completely screened  by the charged gauge field condensate. Different kind vacuum properties have been derived in the literature due to the instability evolution (see \cite{Ambjorn:1988tm, Ambjorn:1989bd} and references therein). This screening behavior was also investigated on a lattice \cite{Antropov:2010tj}. At finite temperature, the situation is different. Here, new dynamical parameters - temperature Debye's and magnetic masses - are generated and act to stabilize the vacuum as in  scalar field theory.  These important parameters can be calculated in perturbation theory with unstable vacuum. The stabilization is expected at high temperature when the values of the temperature mass squared  exceed the value of the tachyonic mass squared. Assuming such a scenario, we can carry out calculations with unstable mode included. In this case we expect that the stabilization of the spectrum will come and external field is not screened and presents  at high temperature. Analytic calculations give a possibility for determining  different parameters constructed  aut of field strength and temperature, derivation of non-analytic relations between these parameters. These calculations also can serve as a background for numeric MC simulations and show possible qualitative picture for the phenomenon studied.
 By joining the  results of  both calculations we obtain a more deep and multilateral notion about it.

First problem to study is space scale of gauge fields at finite temperature in the background field presence. To investigate this problem, the polarization tensor of neutral gauge fields $\Pi(T, \tilde{B}, k_4, \vec{k})$, where $k_4, \vec{k}$ are gluon momentum components, must be calculated. Before doing that, we remind the structure of the gauge fields in the SM. This is necessary, because at high temperature the $SU(2)( SU(3)_c)$ neutral components of gauge fields are generated spontaneously.

\begin{figure}
\begin{center}
\includegraphics[bb=0 0 361 143,width=0.35\textwidth]{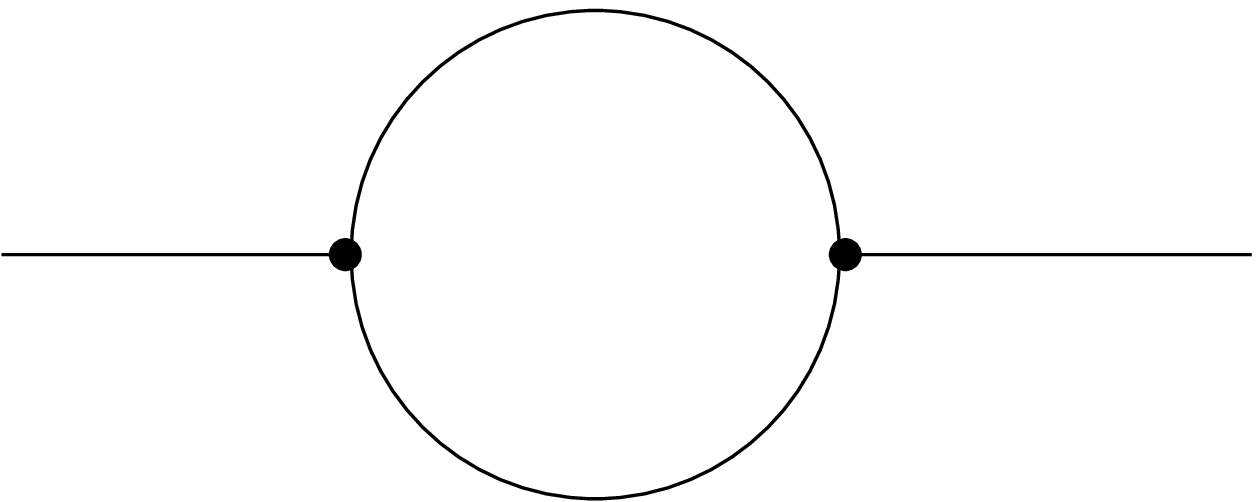}\hskip 2cm
\includegraphics[bb=0 0 217 84,width=0.35\textwidth]{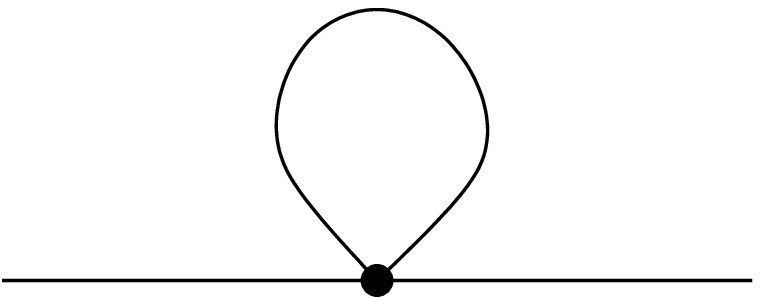}
\end{center}
\caption{Basic graph for polarization tensor (left figure) and graph with
one vertex and a closed line (right figure).} \label{figure:G34}
\end{figure}

The polarization tensor consists basically of the graphs shown in Fig.~\ref{figure:G34} where gluon and ghost lines must be inserted.

The electromagnetic $A_\mu$ and $Z$-boson potentials read
\begin{eqnarray}\label{AZ}
 A_\mu &=& \frac{1}{\sqrt{g^2 + g'^2}} ( g' A^3_\mu + g b_\mu ), \\ \nonumber
 Z_\mu &=& \frac{1}{\sqrt{g^2 + g'^2}} ( g A^3_\mu - g' b_\mu ),
\end{eqnarray}
where $A^3_\mu, b_\mu$ are the Yang-Mills gauge field third projection in the weak isospin space and the potential of the hypercharge gauge field, and $g$ and $g'$ are $SU(2)$ and $U(1)_Y$ couplings, correspondingly. At high temperature, $ A^3_\mu \not = 0$, the field is spontaneously generated. The hypermagnetic field is not spontaneously generated, $ b_\mu = 0.$ After the electroweak phase transition, the $Z$-boson acquires mass and the field $Z_\mu$ is screened. The only component $A_\mu = \frac{1}{\sqrt{g^2 + g'^2}} g' A^3_\mu = \sin \theta_W A^3_\mu$ remains non-zero. Here $\theta_W $ is the Weinberg angle, $\tan \theta_W = \frac{g'}{g}$. Hence, at low temperature the remnant of the field $\tilde{B}$ is the magnetic field $B = \sin\theta_W \tilde{B}$. Below, for brevity, we write $B$ instead $\tilde{B}$ for the background field. The difference will be taken into account when necessary.

 In the field presence  it is reasonable to consider color gluon field, $V^a_\mu$ ($a = 1,2,3 $), separately in terms of charged $W$-bosons (gluons), $ W_\mu^{\pm} = \frac{1}{\sqrt{2}}(V^1_\mu \pm i V^2_\mu), $ and neutral, $V^3_\mu = A^3_\mu$, ones.

The one-loop neutral gluon polarization tensor (PT) in the background field at finite temperature was calculated and investigated in ~\cite{Bordag:2006pr}. An imaginary time formalism was used and, in particular, both the parameters, $m_D(B,T)$ and $m_{magn.}(T,B)$ have been derived.

In general, the PT can be presented  in the form \cite{Bordag:2006pr}:
\begin{equation}\label{exp}
 \Pi_{\lambda\lambda'}(k)=\sum_{i=1}^{10} \ \Pi^{(i)}(k,B,T) \
 T^{(i)}_{\lambda\lambda'} \ ,
\end{equation}
where $T^{(i)}_{\lambda\lambda'}$ are ten structures out of the momentum $k_\mu$, medium velocity $u_\mu$ and $\delta_{\lambda \lambda'}$ and the form factors $\Pi^{(i)}(k)$ depend on the external momentum $k_\mu$ through the variables $l^2 = k_4^2 + k_3^2$ and $h^2 = k_1^2 + k_2^2$ at zero temperature and $h^2, k_4$ and $k_3$ at finite temperature. These structures form a complete tensor basis.

\begin{figure}
\begin{center}
\includegraphics[bb=0 0 398 171,width=0.5\textwidth]{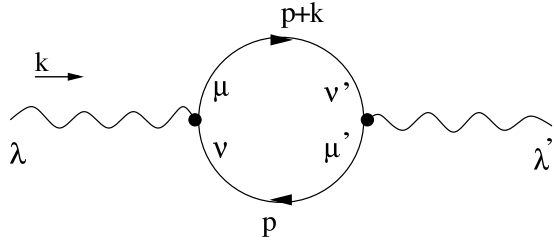}
\end{center}
\caption{The neutral polarization tensor.}\label{figure:Pineu}
\end{figure}

The one-loop PT has the following representation in momentum space (see Fig.~\ref{figure:Pineu})
\begin{equation}\label{NPT}
 \Pi_{\lambda\lambda'}(k)= T \sum\limits_{N=- \infty}^{+\infty}\int\limits_{- \infty}^{+ \infty}\frac{d^3 p}{(2\pi)^3}~\Pi(p,p_4,k,k_4)_{\lambda\lambda'},
\end{equation}
where in the integrand we noted explicitly the dependence on the external momentum and the momentum inside loops. The integrand looks as follows
\begin{eqnarray}\label{NPT1}
 \Pi_{\lambda\lambda'}(p,k)&=&\Gamma_{\mu\nu\lambda}G_{\mu\mu'}(p)\Gamma_{\mu'\nu'\lambda'} G_{\nu'\nu}(p-k)\\
 &&-p_{\lambda}G(p)(p-k)_{\lambda'}G(p-k)-(p-k)_{\lambda}G(p)p_{\lambda'}G(p-k)\nonumber \\
 && + G_{\lambda\lambda'}(p) + G_{\lambda'\lambda}(p) - 2 \delta_{\lambda\lambda'} {\rm Tr}
 G(p).\nonumber
\end{eqnarray}
The second line gives the contribution from the ghost loops and the third one is due to the tadpole diagrams.
The vertex factor is
\begin{equation}\label{Gamma}
 \Gamma_{\mu\nu\lambda}=\delta_{\mu\nu}(k-2p)_\lambda+\delta_{\lambda\mu}(p+k)_\nu+\delta_{\lambda\nu}(p-2k)_\mu.
\end{equation}
It should be remarked that all graphs and combinatorial factors are exactly the same as in the well known case without magnetic field. On this level the only difference is in the meaning of the momentum $p_\mu = i \partial_\mu + i B_\mu$ which in our case depends on the background magnetic field. The four particle vertexes are momentum independent and have the same form as at zero external field.

The propagators are given by
\begin{eqnarray}\label{prop0}
 &&G(p)=\frac{1}{p^2}=\int_0^\infty ds \
 e^{-sp^2}, \nonumber \\ && G(p-k)=\frac{1}{(p-k)^2}=\int_0^\infty dt \
 e^{-t(p-k)^2}
\end{eqnarray}
for the scalar lines and by
\begin{eqnarray}\label{prop}
 G_{\lambda \lambda'}(p)&=&\left(\frac{1}{p^2+2iF}\right)_{\lambda\lambda'}=\int_0^\infty ds \
 e^{-sp^2}E^{-s}_{\lambda\lambda'} ,\nonumber\\
 G_{\lambda \lambda'}(p-k)&=&\left(\frac{1}{(p-k)^2+2iF}\right)_{\lambda\lambda'}=\int_0^\infty dt \ e^{-t(p-k)^2}E^{-t}_{\lambda\lambda'}
\end{eqnarray}
for the vector lines with
\begin{eqnarray}\label{E}
 E^s_{\lambda\lambda'} \equiv
 \left(e^{2isF}\right)_{\lambda\lambda'} = \delta^{||}_{\lambda\lambda'}+iF_{\lambda\lambda'}\sinh(2s)+\delta^\perp_{\lambda\lambda'} \cosh(2s).
\end{eqnarray}

Here and below, for simplicity, we set the field strength $g B = 1$. That means we measure all quantities in units of $g B$. To return to the dimensional  variables one has to substitute $s \to g B s$, etc.

At zero temperature, the momentum integration can be carried out by means of Schwinger's algebraic procedure \cite{Schwinger:1973kp} and converted into an integration over two scalar proper-time para\-meters, $s$ and $t$. Here we adduce the known results in order to present their modifications at $T \not = 0$ \cite{Bordag:2006pr}. The basic exponential is
\begin{equation}\label{mexp}
 \Theta=e^{-sp^2}e^{-t(p-k)^2}
\end{equation}
and the integration over the momentum $p$ is denoted by the average $\langle\dots\rangle$. The following formula holds:
\begin{eqnarray}\label{Theta}
 \langle\Theta\rangle=\frac{\exp\left[-k\left(\frac{st}{s+t}\delta^{||}+\frac{ST}{S+T}\delta^{\perp}\right)
 k \right]}{(4\pi)^2(s+t)\sinh(s+t)}
\end{eqnarray}
with $S=\tanh(s)$ and $T=\tanh(t)$.

Within the developed formalism, the polarization tensor becomes the expression of the type
\begin{equation}\label{Pist}
 \Pi_{\lambda\lambda'}(k)=\int_0^\infty\int_0^\infty \ ds \ dt
 ~ M_{\lambda\lambda'}(p,k) \langle \Theta_T(s,t)\rangle,
\end{equation}
where in $M_{\lambda\lambda'}(p,k)$ we collected all factors appearing from the vertexes and from the lines except for that going into $\Theta_T(s,t)$ (Eq.~(59) in Ref.~\cite{Bordag:2006pr}):
\begin{eqnarray}\label{ThetaT1}
 \langle\Theta (s, t)\rangle_T &=& \sum\limits_{N=-\infty}^{ +\infty} \langle \Theta (s, t) \rangle \ \ \exp\left(-
 \frac{N^2}{4 T^2 (s + t)} + i \frac{k_4 t N}{(s + t) T} \right)\\\nonumber
 &=& \sum\limits_{ N = -\infty}^{ +\infty}\Theta_T (s, t).
\end{eqnarray}
We introduced the notation $\Theta_T(s,t)$ which is the basic function appearing in all form factors. The function $\langle \Theta (s, t) \rangle $ is given in Eq.~(\ref{Theta}). In this way the finite and zero temperature formalisms are related. All the necessary formulas for momentum integration are given in Ref.~\cite{Bordag:2006pr}.

Now, we calculate the magnetic mass of neutral gluons. It can be determined in the imaginary time formalism through the PT,
\begin{equation}\label{mneutral}
 m^2_{magn.}(s) = \langle s | \Pi(k,B,T)|s \rangle_{k_4 = 0, k^2 \to 0}.
\end{equation}
Here, $k_4 = 2 \pi N T$, $N = 0, \pm 1, \pm 2 ,...$ is a Matsubara frequency, $k^2 = l^2 + h^2 $ is a momentum squared. The mean value is calculated in the two states of polarization (denoted as $s = 1$ and $s = 2$ \cite{Bordag:2005br, Bordag:2006pr}) transverse to the gluon momentum $k_\mu$.

The factors $M^{(i)} $ giving contributions to the matrix elements of $\Pi$ for the infrared limit of interest in Eq.~(\ref{mneutral}) to the states $s = 1$, $s = 2$ are \cite{Bordag:2006pr}:
\begin{eqnarray}\label{Mia}
 M_2&=&4 \frac{1 - \cosh(q) \cosh(\xi)}{(\sinh(q))^2} - 2 + 8 \cosh(q)\cosh(\xi), \nonumber \\
 M_3&=&- 2\cosh(2q) \frac{\xi\sinh(\xi)}{q\sinh(q)} - 2 + 6\cosh(\xi)\cosh(q), \nonumber \\
 M_5&=& - 2 + 2 \cosh(q)\cosh(\xi),
\end{eqnarray}
where $q = s + t$, $\xi = s - t = q ( 2 u - 1)$ and $s = q u$, $t = q(1 - u)$. The contributions to the particular polarization states $s$ are (Eq.~(142) Ref.~\cite{Bordag:2006pr}):
\begin{eqnarray}\label{Pitr}
 \langle s=1| \Pi(k) |s=1\rangle &=& h^2 \Pi_2,\nonumber \\
 \langle s=2| \Pi(k) |s=2\rangle &=& h^2 \left( \Pi_3 + \Pi_5 \right).
\end{eqnarray}
We have to calculate the form factors $\Pi_2$, $\Pi_3$ and $\Pi_5$.

These matrix elements are the product of $h^2$ and expressions which have a finite limit for $h^2=0$,
\begin{equation}\label{2h2}
 \Pi_i=\Pi^{(0)}_i+O(h^2).
\end{equation}
The quantities $\Pi^{(0)}_i$ can be calculated analytically in terms of Riemann  Zeta-function.

From the above expressions (\ref{Pist})-(\ref{Mia}), in leading order for $T\to\infty$, which picks just the $N=0$-contribution, we note
\begin{equation}\label{ffi}
 \Pi^{(0)}_i =\frac{g^2}{(4\pi)^{3/2}}\frac{T}{\sqrt{g B}}
 \int\limits_0^1 d u \int\limits_0^{\infty}
 \frac{d q \sqrt{q}}{\sinh(q)}\ M_i(q,u).
\end{equation}
In these expressions, the integration over $u$ can be carried out explicitly,
\begin{equation}\label{ffii}
\Pi^{(0)}_i =\frac{g^2}{(4\pi)^{3/2}}\frac{T}{\sqrt{g B}}
 \int\limits_0^{\infty} \frac{d q \sqrt{q}}{\sinh(q)}\ M_i(q)
\end{equation}
with
\begin{eqnarray}\label{2M}
 M_2(q)&=&-2-\frac{4}{q}\coth(q)+\frac{4}{\sinh(q)^2}+\frac{4}{q} \sinh(2q),\nonumber\\
 M_3(q)&=&-2-\frac{2}{q^2}\cosh(2q)\left(-1+q\coth(q)\right)+\frac{3}{q} \sinh(2q),\nonumber\\
 M_5(q)&=&-2+\frac{1}{q}\sinh(2q),
\end{eqnarray}
and the $q$-integrations remain.

These expressions are formally divergent for $q\to\infty$. This divergence results from the tachyonic mode. Here we have to remember that all formulas above are written in Euclidean representation (basically, for technical reasons). In fact, we have to start from the Minkowski space representation which can be reached by an 'Anti'-Wick rotation, $q\to q e^{i\pi/2}$. In the Minkowski space representation the parametric integrals are convergent using the usual '$i\epsilon$'- prescription. But then, the contribution from the tachyonic mode in the loop cannot be Wick-rotated since, in momentum space, the corresponding pole is on the 'wrong' side of the imaginary axis of the momentum $p_0$. However, it can be 'Anti'-Wick rotated delivering a exponentially fast converging integral. The remaining part can be Wick rotated as usual. In this way, if starting from the Euclidean representation, the tachyonic part must be 'Anti'-Wick rotated twice, $q\to q e^{i\pi}$. The remaining part can be kept as it is. The subdivision into tachyonic and remaining parts must be done according to the behavior for $q\to\infty$. There is a freedom left of redistribution power like contributions. It can be used to avoid singularities in $q=0$.

Performing all the calculations we get the corresponding numerical values \cite{Antropov:2010tj, Bordag:2006pr},
\begin{eqnarray}\label{2tp3}
 \Pi^{(0)}_2&=&\frac{g^2}{(4\pi)^{3/2}}\frac{T}{\sqrt{g B}}\left(-5.80+7.09 i\right),\nonumber\\
 \Pi^{(0)}_3&=&\frac{g^2}{(4\pi)^{3/2}}\frac{T}{\sqrt{g B}}\left(1.04-8.9 i\right),\nonumber\\
 \Pi^{(0)}_5&=&\frac{g^2}{(4\pi)^{3/2}}\frac{T}{\sqrt{g B}}\left(-4.21+1.8 i\right)\,.
\end{eqnarray}

The above expressions have to be used in Eq.~(\ref{Pitr}) to obtain final result. The sum of $\Pi_3 + \Pi_5 $ equals, $\Pi_3 + \Pi_5 = \left[-3.17 - 7.09 i \right]$. The imaginary part is signaling the instability of the state because of the tachyon mode, and the real one is responsible for the screening of transverse gluon fields. The real and imaginary parts are of the same order of magnitude. This is similar to the case of Landau's damping at finite temperature.

Let us turn to the real part and substitute it in the operator Schwinger-Dyson equation
\begin{equation}\label{SDe}
 D^{-1}(k^2) = k^2 - \Pi(k)
\end{equation}
for the neutral gluon Green function. We obtain for the mean values
\begin{eqnarray}\label{SDes1}
 \langle ~s=1~|D^{-1}(h^2)|~s=1 ~\rangle &=& h^2 - {\rm Re} ( \Pi_2) ~h^2\\\nonumber
 &=& h^2 \left( 1 + 5.8 \frac{T}{\sqrt{g B}} \right)
\end{eqnarray}
and
\begin{eqnarray}\label{SDes2}
 \langle~ s=2~|D^{-1}(h^2)|~s=2~ \rangle &=& h^2 - {\rm Re} ( \Pi_3 + \Pi_5) ~h^2 \\\nonumber
 &=& h^2 \left( 1 + 7.09 \frac{T}{\sqrt{g B}} \right).
\end{eqnarray}
These are the expressions of interest.

Two important conclusions follow from Eqs.~(\ref{SDes1})-(\ref{SDes2}). First, for the transverse modes in the field presence, there is no fictitious pole similar to that of in the one-loop approximation for zero external field background at finite temperature \cite{Kalashnikov:1982sc, Landsman:1986uw, Kraemmer:2003gd}. The external field acts as some kind resummation removing this singularity. Second, there is no magnetic screening mass in one-loop order. The transverse components of the gluon field remain long-range in this approximation, as at zero external field \cite{Kalashnikov:1982sc}.

Possible resolutions of the zero one-loop magnetic mass are obvious: 1) the mass is generated in some kind resummation of perturbation series (as this is well known at zero external field case \cite{Hidaka:2009hs, Braaten:1989mz}); 2) there are no magnetic mass for neutral gluons as in the case of usual magnetic fields. The problem requires non-perturbative methods of computation.

Second characteristic to calculate is the Debye mass determined in a standard way: $m^2_D = -\Pi_{44}(T,B,$ $k_4=0,\vec{k}\to 0)$ \cite{Kalashnikov:1982sc}. It reads \cite{Bordag:2006pr}
\begin{eqnarray}\label{mDB}
 m_D^2(B) = \frac{1}{4\pi^2} \int\limits_0^{\infty}\frac{dq}{q}\sum\limits_{N=1}^{+\infty}
 \frac{N^2}{ qT^2} ~ \frac{B \cosh(2Bq)}{ \sinh(Bq)}e^{-\frac{N^2}{4qT^2}}
 \equiv \frac{2}{3}T^2 \ f\left(\frac{B}{4 T^2}\right),
\end{eqnarray}
where the dimensional   parameters are restored. The function $f(s)$ is dimensionless and it depends on the dimensionless variable $s = \frac{B}{4 T^2}$. It is chosen to satisfy $f(0) = 1$ such that it describes just the change which comes in from the magnetic field.

The function $f(s)$ can be easily computed numerically. Also, its asymptotic expansion is easy to obtain. We consider small $s$, i.e., high temperature, and represent $f(s)$ in the form
\begin{equation}\label{fs}
 f(s) = \frac{6}{\pi^2} s^2 \int\limits_0^{\infty}\frac{dq}{q^2}\sum\limits_{N=1}^{+\infty} N^2
 \left[\frac{1}{q} + \frac{ \cosh(2q)}{ \sinh( q)} - \frac{1}{q}\right]~ e^{-\frac{N^2s}{q}}\ .
\end{equation}
Here the first term in the square brackets delivers the zero field limit, $f(0) = 1$.

The detailed calculation of it are carried out in Ref.~\cite{Bordag:2006pr}, for the function $f(s) $ we obtain
\begin{eqnarray}\label{fsr}
 f(s) = 1 + \left[\frac{3}{\sqrt{2}\pi}\left(\sqrt{2}-1\right) \zeta\left(\frac{1}{2}\right) -
 \frac{3}{2\pi} \right] \sqrt{ s} + \frac{25}{4}
 \frac{\zeta(3)}{\pi^4} s^2 - i \frac{3}{2\pi} \sqrt{s}.
\end{eqnarray}
Hence, for the Debye mass we derive
\begin{eqnarray}\label{mDBfin}
 m^2_D(B) &&= \frac{2}{3}g^2 T^2 \left[1 - 0.8859\left(\frac{\sqrt{g B}}{2T}\right) + 0.4775 \left(\frac{g^2 B^2}{16 T^4}\right) \right.\nonumber \\
 &&\left. - i~ 0.4775 \left(\frac{\sqrt{B}}{2T}\right) + O\left(\frac{g^3 B^3}{T^6}\right)\right],
\end{eqnarray}
where the general factor $g^2$ and numeric values of the coefficients in Eq.~(\ref{fsr}) were substituted.

In this way, from Eq.~(\ref{fs}) the Debye mass can be determined in terms of the Riemann Zeta-function. As expected, it has an imaginary part which results from the tachyonic instability. We note that the Debye mass in the magnetic field is smaller than without it. That is, in the field presence a color electric field becomes a more long-range one compared to the zero field case. From Eq.~(\ref{mDBfin}) it follows also that the imaginary part is next-to-leading order compared to the real one. The ratio $\rho = {\rm Im} ~m^2_D(B)/{\rm Re}~ m^2_D(B) \ll 1$. This property, in particular, means that that perturbative methods of calculations are sufficient for determining this parameter. In contrast, for the magnetic mass case the stability parameter $\rho ={\rm Im} \langle\Pi(B)\rangle/{\rm Re} \langle\Pi(B)\rangle $ is of the order $\sim 1$. In the latter case, $\rho$ has the order of usual Landau's damping parameter in plasma.

Similar behavior is proper to the neutral gluon field in the $SU(3)_c$ case.

\section{Polarization tensor of charged gauge fields at finite temperature}
Let us investigate the PT of charged gluons ($W$-bosons) in $SU(2)$ gluodynamics in the background field (\ref{potentialB}) at finite temperature \cite{Bordag:2008wp, Bordag:2012tk}.
In the field, the polarization tensor can be constructed out of the vectors $l_\mu$, $h_\mu$ and $d_\mu$,
\begin{eqnarray}\label{vecp}
 &&l_\mu=\left(\begin{array}{c}0\\0\\p_3\\p_4\end{array}\right),~
 h_\mu=\left(\begin{array}{c}p_1\\p_2\\0\\0\end{array}\right), ~ d_\mu=\left(\begin{array}{c}p_2\\-p_1\\0\\0\end{array}\right),
\end{eqnarray}
where the third vector is $d_\mu= F_{\mu\nu}p_\nu$ and we note $p_\lambda=l_\lambda+h_\lambda$, and the matrixes
\begin{equation}\label{F}
 \delta^{||}_{\mu\lambda}=\left(\begin{array}{cccc}0&0&0&0\\0&0&0&0\\0&0&1&0
 \\0&0&0&1\end{array}
 \right), \
 \delta^{\perp}_{\mu\lambda}=\left(\begin{array}{cccc}1&0&0&0\\0&1&0&0\\0&0&0&0
 \\0&0&0&0\end{array}
 \right), \
 F_{\mu\lambda}=\left(\begin{array}{cccc}0&1&0&0\\-1&0&0&0\\0&0&0&0\\0&0&0&0\end{array}
 \right).
\end{equation}
Here, $ p_\lambda$ is a momentum in the external magnetic field. It is an operator and it obeys the known commutation relation
\begin{equation}\label{commrelp}
 [{p}_\mu,{p}_\nu ]=iF_{\mu\nu}.
\end{equation}
There are no more vectors or constant matrices, the polarization tensor may depend on. For instance, this reflects the residual rotational symmetry in the ($p_3$,$p_4$)-subspace. Hence the operator structures can also be constructed out of these quantities only. We mention that with our choice $B=1$, $F_{\mu\nu}$ in (\ref{F}) is the field strength of the background field.

Before writing down the decomposition of the PT in terms of form factors we mention one property which follows directly from the basic commutator relation the momentum $p_\mu$ obeys, namely $p_\lambda p^2=(p^2\delta_{\lambda\lambda'}+2iF_{\lambda\lambda'})p_{\lambda'}$. As a consequence, for a function of $p^2$ the relations
\begin{eqnarray}\label{p2+2iF}
 p_\lambda f(p^2)&=&f(p^2+2iF)_{\lambda\lambda'}p_{\lambda'} \ ,\nonumber \\
 f(p^2)p_\lambda &=&p_{\lambda'}f(p^2+2iF)_{\lambda'\lambda}
\end{eqnarray}
hold where now $f$ must be viewed as a function of a matrix so that it itself becomes a matrix carrying the indices $\lambda$ and $\lambda'$. The same is true with $h^2$ in place of $p^2$.

\begin{figure}
\begin{center}
\includegraphics[bb=0 0 378 257,width=0.5\textwidth]{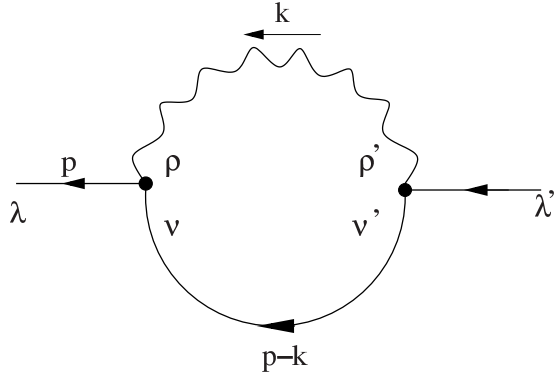}
\end{center}
\caption{The charged polarization tensor.}\label{figure:Picha}
\end{figure}

The decomposition of the polarization tensor can be written in the form (see Fig.~\ref{figure:Picha})
\begin{equation}\label{expp}
 \Pi_{\lambda\lambda'}(p)=\sum_{i}\ \Pi^{(i)}(l^2,h^2+2iF)_{\lambda\lambda''} \ T^{(i)}_{\lambda''\lambda'}
 +\Pi^{\rm D} T^{\rm D}_{\lambda\lambda'}.
\end{equation}
In general, the sum includes ten structures $T^{(i)}_{\lambda\lambda'}$, and $T^{\rm D}_{\lambda\lambda'} = u_\lambda u_\lambda'$
defined in ~\cite{Bordag:2005br, Bordag:2008wp}. The form factors
$\Pi^{(i)}(l^2,h^2+2iF)_{\lambda\lambda'}$ depend on
$l^2$ and $h^2$ only (besides their dependence on the matrices in (\ref{F})). In (\ref{expp}) the form factors can be placed also on the right from the operator structures applying both relations (\ref{p2+2iF}).

The calculations of the charged PT differ in an essential way from the neutral case although all the steps and methods are similar. The main difference consists in the dependence of form factors on the operators $h^2+2iF$. We use the representation of the polarization tensor given in \cite{Bordag:2008wp}. In momentum representation, the initial expression reads
\begin{eqnarray}\label{Pi}
 \Pi_{\lambda\lambda'}(p)&=&\int\frac{dk}{(2\pi)^4} \
 \biggl\{\Gamma_{\lambda\nu\rho}G_{\nu\nu'}(p-k)\Gamma_{\lambda'\nu'\rho'}G_{\rho\rho'}(k)\nonumber \\
 &&+(p-k)_{\lambda}G(p-k)k_{\lambda'}G(k)+k_{\lambda}G(p-k)(p-k)_{\lambda'}G(k)\biggr\}\nonumber\\
 [6pt]&&+\Pi^{\rm tadpol}_{\lambda\lambda'} \ ,
\end{eqnarray}
where the second line results from the ghost contribution and the tadpole contribution is given by
\begin{eqnarray}\label{Tp}
 \Pi^{\rm tadpol}_{\lambda\lambda'}&=&
 \int\frac{dk}{(2\pi)^4} \ \biggl\{\delta_{\lambda\lambda'}G_{\rho\rho}(k)-G(k)_{\lambda\lambda'}\biggr\} \nonumber \\ &&
 +\int\frac{dp}{(2\pi)^4} \ \biggl\{\delta_{\lambda\lambda'}G_{\rho\rho}(p)+G_{\lambda'\lambda}(p)-2G_{\lambda\lambda'}(p)\biggr\}.
\end{eqnarray}
The vertex factor and propagators are given in Eqs.~(\ref{Gamma}) - (\ref{prop}).

To proceed further, it is convenient to  turn  the momentum and the gluon field in color space into the charged basis,
\begin{equation}\label{Bma}
 p_\mu=B_{\mu\alpha}\ p_\alpha \quad \mbox{with} \quad B_{\mu\alpha}=\left(\begin{array}{cc}\frac{1}{\sqrt{2}}
 \left(\begin{array}{cc}1&1\\i&-i\end{array}\right) & 0\\ 0 & \left(\begin{array}{cc}1&0\\0&1\end{array}\right) \end{array}\right)_{\mu\alpha} ,
\end{equation}
we have a color neutral and a color charged field, both of spin 1. The charged field will occupy in the presence of the background field Landau's levels and we expand with respect to the corresponding eigenfunctions. In this space, the momentum $p_\mu$ of the charged gluon is an operator, whose components fulfill the commutation relation
\begin{equation}\label{com}
 [p_\alpha,p_\beta]=iF_{\alpha\beta}\equiv i\left(
 \begin{array}{cccc}
  -1&0&0&0\\0&1&0&0\\0&0&1&0\\0&0&0&1
 \end{array}
 \right)_{\alpha\beta}B.
\end{equation}
As before, in this expression we write $B$ instead $g B$. In what follows,  we also put $B = 1$, for short.

A basis in this space is given by the vectors $\mid n,\sigma\rangle_\mu$, Eq.~(30) in \cite{Bordag:2005br}. The tree level energies of these states are
\begin{equation}\label{En}
 E_n=l_4^2+l_3^2+B(2n+1+2\sigma)\qquad(n=0,1,\dots, ~~\sigma=\pm 1),
\end{equation}
where $l_3$ and $l_4$ are the momenta in parallel to the background field and imaginary time, respectively. The lower state is $n=0$ with $\sigma=-1$. It is tachyonic one. In this section, in contrast to Eq.~(\ref{spectrum}), we use other system for marking the charged gluon states in the field. In the charged basis, the lower ({\it tachyonic}) state is
\begin{equation}\label{t}
 \mid t\rangle_\alpha\equiv \mid 0,-1\rangle_\alpha
 =\left(\begin{array}{c}1\\0\\0\\0\end{array}\right)_\alpha\mid 0\rangle ,
\end{equation}
where $\mid 0\rangle$ is the lower Landau level which is annihilated by the operator $a$ in
\begin{equation}\label{pal}
 p_\alpha=\left(\begin{array}{c}ia^\dagger\\-ia\\l_3\\l_4\end{array}\right)_\alpha
\end{equation}
and we note
\begin{equation}\label{pmual}
 p_\mu\mid t\rangle_\mu=p^\dagger_\alpha\mid t\rangle_\alpha=0.
\end{equation}
Below, we will also use the notations $l^2=l_3^2+l_4^2$ and $h^2=p_1^2+p_2^2=aa^\dagger+a^\dagger a$.

Now, we continue with using the representation of the PT as given by Eq.~(51) in \cite{Bordag:2008wp}. It results from the proper time representation of the propagators and the vertex factor given in Eqs.~(\ref{prop0}), (\ref{prop}), (\ref{Gamma}) and integration over $k$ in Eq.~(\ref{Pi}).

The one-loop PT in the imaginary time formalism can be written in the form of the integral over two proper time parameters and temperature sum
\begin{equation}\label{Pi2}
 \Pi_{\lambda\lambda'}=
 \sum_N \int dsdt \ \langle {\Theta_T}\left[ \sum_{i,j}\hat{M}^{i,j}_{\lambda\lambda'}+
  \hat{M}^{\rm gh}_{\lambda\lambda'}\right]\rangle
 + \ \Pi^{\rm tadpol}_{\lambda\lambda'}
\end{equation}
with $\hat{M}^{i,j}_{\lambda\lambda'}$ coming from the main gluon one-loop diagram, $\hat{M}^{\rm gh}_{\lambda\lambda'}$ is the corresponding contribution from the ghost loop.
The last term presents the contributions of the tadpole diagrams. Here we introduced the basic average
\begin{equation}\label{ThetaT}
 \Theta_T(l^2,h^2)= \ \exp\left\{-\frac{N^2}{4(s+t)T^2}+2s(\tilde{u}p)\right\}\Theta(s,t)\ ,
\end{equation}
where $\Theta(s,t)$ is given by
\begin{equation}\label{Pitp}
 \Theta(s,t)=\frac{\exp(-H)}{(4\pi)^2(s+t)\sqrt{\Delta}},
\end{equation}
and
\begin{equation}\label{uti}
 \tilde{u}_\lambda=\frac{iN}{2(s+t)T}\ u_\lambda.
\end{equation}
This average is what comes at finite temperature in place of (\ref{Pitp}),
\begin{equation}\label{AvT}
 T\sum_{N = -\infty}^\infty \int\frac{d^3k}{(2\pi)^3} \
 = \langle \hat{\Theta}\rangle_T =\Theta_T(l^2,h^2) \ .
\end{equation}
The following notations are used \cite{Bordag:2012tk}:
\begin{eqnarray}\label{nota}
 H&=&\frac{st}{s+t}\, l^2+m(s,t)h^2\,, \nonumber\\
 m(s,t)&=&s+\frac12\ln\frac{\mu_-}{\mu_+}\,, \nonumber\\
 \Delta&=&\mu_- \, \mu_+ ,\nonumber\\
 \mu_\pm&=&t+\sinh(s)e^{\pm s}.
\end{eqnarray}

The details on calculations of the PT (\ref{expp}), (\ref{Pi2}) are given in the mentioned papers. We note here that the PT is not transversal. Instead, the following weaker property holds: $p_\lambda \Pi_{\lambda \lambda'} p_\lambda' = 0$. The important technical achievement of these calculations is the carried out integration by parts, what results in the simple explicit expressions for the form factors.

In the present study, we are interested in two special cases - the Debye, $m_D^{W},$ and magnetic, $m_{magn.}^{W},$ masses for charged gluons. As concerns the Debye mass, it coincides with Eqs.~(\ref{mDB}), (\ref{mDBfin}) obtained above for the neutral PT. It obviously must hold in the limit of zero background field. However, that it holds for any magnetic field strength is to some extend unexpected result.

Now, we consider the PT in the limit of high temperature, $T\to\infty$. It is obtained by taking the contribution of the zeroth Matsubara frequency, $N=0$ in the otherwise unchanged expression. This is, of course, the known dimensional reduction to a theory without temperature in three dimensions. In the representation in terms of parametric integrals, the tachyonic projection of the polarization tensor takes the form \cite{Bordag:2012tk}:
\begin{eqnarray}\label{PiT}
 \langle t\mid \Pi\mid t\rangle &\stackrel{T\to\infty}{\sim}&
 \frac{T\sqrt{B}}{(4\pi)^{3/2}} \Bigg\{
 \int\frac{ds\, dt}{\sqrt{s+t}}
 \left(\frac{4}{\mu_-}+4\frac{s+te^{2s}}{s+t}\frac{l^2}{B}\right)\ \frac{e^{-\frac{st}{s+t}\frac{l^2}{B}-s}}{\mu_-}\nonumber\\
 [5pt]&&+\int_0^\infty\frac{dq}{\sqrt{q}}
 \left(\frac{-2}{q}-\frac{2}{\sinh(q)}-4\cosh(q)\right) \Bigg\},
\end{eqnarray}
where now $l^2=l_3^2$. In this expression, we removed the regularization. This is possible since the (\ref{PiT}) does not contain ultraviolet divergencies. The linear ones cancel as in four dimensions and the logarithmic ones do not appear due to the dimensionality.

In the following we focus on the magnetic mass of the charged gluon,
\begin{equation}\label{mch}
 m_{\rm magn.}^2 = - \langle t| \Pi(B, T, p_4 = 0, p_3 \to 0)| t\rangle.
\end{equation}
It is given by Eq.~(\ref{PiT}) with $l^2=0$ in the tachyonic projection. So it remains to treat the tachyonic mode in (\ref{PiT}). Because of $l^2=0$ this is here simpler than in the preceding section. Accounting for the changes dimensionality it can be calculated easily and it delivers the imaginary part,
\begin{equation}\label{rmass}
 {\rm Im}(\langle t\mid \Pi\mid t\rangle_{T\to\infty})=-2i\sqrt{\pi}~g^2 \frac{T\sqrt{g B}}{(4\pi)^{3/2}}.
\end{equation}
The real part is also much simpler than in the preceding section and it reads
\begin{eqnarray}\label{mmass}
 {\rm Re}( \langle t\mid \Pi\mid t\rangle_{T\to\infty}) &=& g^2~ \frac{T\sqrt{g B}}{(4\pi)^{3/2}}
 \int_0^\infty\frac{dq}{\sqrt{q}}
 \left[ \int_0^1 du\ \frac{4qe^{-s}}{\mu_-^2}-\frac{2}{q}-\frac{2}{\sinh(q)}-2e^{-q}\right]\nonumber\\
 &\simeq&-3.26 ~g^2 ~\frac{T\sqrt{g B}}{(4\pi)^{3/2}}.
\end{eqnarray}
The last line is a result of numerical integration.

The main result of this calculation is that the ground state projection of the PT is proportional to $T \sqrt{g B}$ allowing for the conclusion that there is a field dependent magnetic mass. Further, it is very important that the real part has a negative sign. Hence it follows that radiation corrections act to stabilize the spectrum of charged gluons at high temperature. The stability parameter is $\rho \sim 1$ as for the neutral gluon case.

In the electroweak sector of the SM, in the restored phase it is common to consider properties of the scalar and gauge fields separately. Here, we follow this tradition and say about $SU(2)$ gluons.

In this and previous section we determined the main screening characteristics of gauge fields at high temperature - Debye's and magnetic masses. We observed that in the field presence these parameters differ in an essential way from that of zero field case. Especially interesting for us here is magnetic mass responsible for properties of magnetic fields at high temperature. So, to obtain the results independent of perturbative expansions, in the next two sections we investigate the spontaneous magnetization and the neutral gluon magnetic mass by considering gauge field theory on the lattice.

\section{Spontaneous vacuum magnetization in lattice investigations}
Now, we are going to investigate the spontaneous vacuum magnetization at finite temperature in $SU(2)$ gluodynamics on the lattice by using the Monte Carlo (MC) simulations. Recent review on the lattice QCD in background fields is given in \cite{D'Elia:2012tr}. In particular, influence of external chromomagnetic fields on the deconfinement phase transition was invested in \cite{Campanelli:2007qn, Cea:2007yv}. It was observed that the phase transition temperature, $T_d$, is strongly dependent on the strength of the field applied.

In our problem \cite{Demchik:2008zz, Demchik:2007ct}, the main object is a magnetic flux instead of a field strength. This is because on the lattice the magnetic field strength is quantized. So, it is very difficult to tune the temperature parameter which corresponds to the field strength generated inside of the whole lattice. To overcome this difficulty, we relate free energy density of the continuous magnetic flux to the effective action according to the definition,
\begin{eqnarray}\label{enactions}
 F(\varphi)=\bar{S}(\varphi, \beta)-\bar{S}(0, \beta),
\end{eqnarray}
where $\bar{S}(\varphi, \beta)$ and $\bar{S}(0, \beta)$ are the effective lattice actions with and without chromomagnetic field, correspondingly, $\varphi$ is the field flux. To detect the spontaneous creation of the field it is necessary to show that free energy has a global minimum at non-zero flux, $\varphi_{min}\not =0$, and its value  is negative in the minimum. Let us discuss this in more details.

As is known, on the lattice free energy is not measured straightforwardly. Instead one measures its derivative with respect to the inverse temperature parameter $\beta$ and then free energy is calculated after corresponding integrations. In our problem, we, first of all, are searching for the qualitative features of the effect. The variable (\ref{enactions}) is the corresponding parameter, which has to be measured at a given temperature. As we showed above, the field strength generated spontaneously at finite temperature is proportional to coupling constant and therefore not strong compared to the squared of the temperature. So, it is expected that variable $F(\varphi)$ takes not large values. Hence, the influence of the flux presence on the parameters of the MC lattice calculation in not essential and can be neglected in what follows. A detailed investigation of the influence of magnetic fields on the MC procedures and parameters is given in \cite{Bali:2011qj, Bali:2012jv}, where, in particular, it is shown that this is really the case. In the course of the standard calculation procedures, formula (\ref{enactions}) can be approximated as,
\begin{eqnarray}\label{enaction1}
 F(\varphi)&=&\int\limits_0^\beta \big( \frac{\partial \bar{S}(\varphi, \beta')}{\partial \beta'} - \frac{\partial \bar{S}( \beta' )}{\partial \beta'}\big)d \beta' \\ \nonumber
 &=& c [\bigl(\bar{S}(\varphi, \beta)-\bar{S}(0, \beta)\bigr) - \bigl(\bar{S}(\varphi, 0 )-\bar{S}(0, 0)\bigr)],
\end{eqnarray}
where $c$ is a positive constant. It is determined by the ratio of the given $\beta$  and   discrete interval $\Delta \beta$ chosen in the approximation.  The  terms in the first brackets  give just $c  F(\varphi)$, and the second brackets is expected to be zero. This is because at zero temperature we have a confinement, and the calculated action with and without  flux is the same.

In the MC simulations, we use the hypercubic lattice $L_t\times L_s^3$ ($L_t<L_s$) with the hypertorus geometry. $L_t$ and $L_s$ are the temporal and the spatial sizes of the lattice, respectively. In the limit of $L_s \to \infty$ the temporal size $L_t$ is related to physical temperature.

The standard Wilson action of the $SU(2)$ lattice gauge theory is
\begin{eqnarray}\label{Action}
 S_W=\beta\sum_x\sum_{\mu<\nu}\left[1-\frac{1}{2}{\rm Tr}
 \left[{\bf U}_\mu(x){\bf U}_\nu(x+a\hat{\mu}){\bf U}^\dag_\mu(x+a\hat{\nu}){\bf U}^\dag_\nu(x)\right]\right],
\end{eqnarray}
where $\beta=4/g^2$ is the lattice coupling constant, $g$ is a bare gauge coupling, ${\bf U}_\mu(x)$ is the link variable located on the link leaving the lattice site $x$ in the $\mu$-th direction. The link variables ${\bf U}_\mu(x)$ are $SU(2)$ matrices decomposed in terms of the unity, $I$, and Pauli $\tau_j$, matrices in color space,
\begin{eqnarray}\label{SU2decomp}
 U_\mu(x)=IU_\mu^0(x)+i\tau_j U_\mu^j(x)=\left(\begin{array}{cc}
 \:\:\:\, U_\mu^0(x)+i U_\mu^3(x) &  U_\mu^2(x)+i U_\mu^1(x)\\
 -U_\mu^2(x)+i U_\mu^1(x) &  U_\mu^0(x)-i U_\mu^3(x)
 \end{array}\right).
\end{eqnarray}

\begin{figure}
\begin{center}
\includegraphics[bb=0 0 259 252,width=0.35\textwidth]{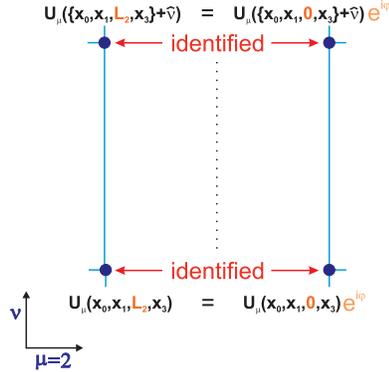}
\end{center}\label{Fig:1}
\caption{The plaquette presentation of the twisted boundary
conditions.}
\end{figure}

Next let us incorporate the external Abelian magnetic field (\ref{potentialB}) into this formalism \cite{Demchik:2008zz, Demchik:2006qj}. The constant homogeneous external flux $\varphi$ in the third spatial direction can be introduced by applying the following twisted boundary conditions (t.b.c.) \cite{Demchik:2006qj}:
\begin{eqnarray}\label{r22}
 & &U_\mu(L_t,x_1,x_2,x_3)=U_\mu(0,x_1,x_2,x_3),\\\nonumber
 & &U_\mu(x_0,L_s,x_2,x_3)=U_\mu(x_0,0,x_2,x_3),\\\nonumber
 & &U_\mu(x_0,x_1,L_s,x_3)=e^{i\varphi} U_\mu(x_0,x_1,0,x_3),\\\nonumber
 & &U_\mu(x_0,x_1,x_2,L_s)=U_\mu(x_0,x_1,x_2,0).
\end{eqnarray}
These give
\begin{eqnarray}\nonumber
&&U_\mu^0(x)= \begin{cases}{\begin{array}{cc}
 U_\mu^0(x)\cos(\varphi)-U_\mu^3(x)\sin(\varphi) & $for $x=(x_0,x_1,L_s,x_3) $ and $ \mu=2 \\
 \!\!\!\!\!\!\!\!\!\!\!\!\!\!\!\!\!\!\!\!\!\!\!\!\!\!\!\!\!\!\!\!\!\!\!\!\!\!\!\!\!\!\!\!\!\!\!\!\!\!\!\!\!\!\!\!U_\mu^0(x) &
 \!\!\!\!\!\!\!\!\!\!\!\!\!\!\!\!\!\!\!\!\!\!\!\!\!\!\!\!\!\!\!\!\!\!\!\!\!\!\!\!\!\!\!\!\!\!\!$for other links$\\
\end{array},
}\end{cases} \\ \nonumber &&U_\mu^3(x)= \begin{cases}{\begin{array}{cc}
 U_\mu^0(x)\sin(\varphi)+U_\mu^3(x)\cos(\varphi) & $for $x=(x_0,x_1,L_s,x_3) $ and $ \mu=2 \\
 \!\!\!\!\!\!\!\!\!\!\!\!\!\!\!\!\!\!\!\!\!\!\!\!\!\!\!\!\!\!\!\!\!\!\!\!\!\!\!\!\!\!\!\!\!\!\!\!\!\!\!\!\!\!\!\!U_\mu^0(x) &
 \!\!\!\!\!\!\!\!\!\!\!\!\!\!\!\!\!\!\!\!\!\!\!\!\!\!\!\!\!\!\!\!\!\!\!\!\!\!\!\!\!\!\!\!\!\!\!$for other links$ \\
\end{array}.
}\end{cases} \\ \nonumber
\end{eqnarray}
The edge links in all directions are identified as usual periodic boundary conditions except for the links in the second spatial direction for which the additional phase $\varphi$
is added (Fig.~\ref{Fig:1}). In the continuum limit, such t.b.c. settle the magnetic field with the potential $\bar{A}_\mu = (0,0, B x^1,0)$ (\ref{potentialB}). The magnetic flux $\varphi$ is measured in angular units and can take continuous values from $0$ to $2\pi$. More details on the t.b.c. can be found in Ref.~\cite{DeGrand:1981yx}.

The MC simulations are carried out by means of the heat bath method. The lattices $2\times 8^3$, $2\times 16^3$ and $4\times 8^3$ at $\beta=3.0$, $5.0$ are considered. These values of the coupling constant correspond to the deconfinement phase and perturbative regime.

The effective action depends smoothly on the flux $\varphi$ in the region $\varphi\sim 0$.\\ So, the free energy density can be fitted by a quadratic function of $\varphi$,
\begin{eqnarray}\label{ffit}
 F(\varphi)=F_{min}+b(\varphi-\varphi_{min})^2.
\end{eqnarray}

In Eq.~(\ref{ffit}), there are three unknown parameters, $F_{min}$, $b$ and $\varphi_{min}$. $\varphi_{min}$ denotes the minimum position of free energy, whereas the $F_{min}$ and $b$ are the free energy density at the minimum and the curvature of the free energy function, correspondingly. They have been fitted by a standard $\chi^2$ method.

\begin{figure}
\begin{center}
\includegraphics[bb=0 0 580 351,width=0.45\textwidth]{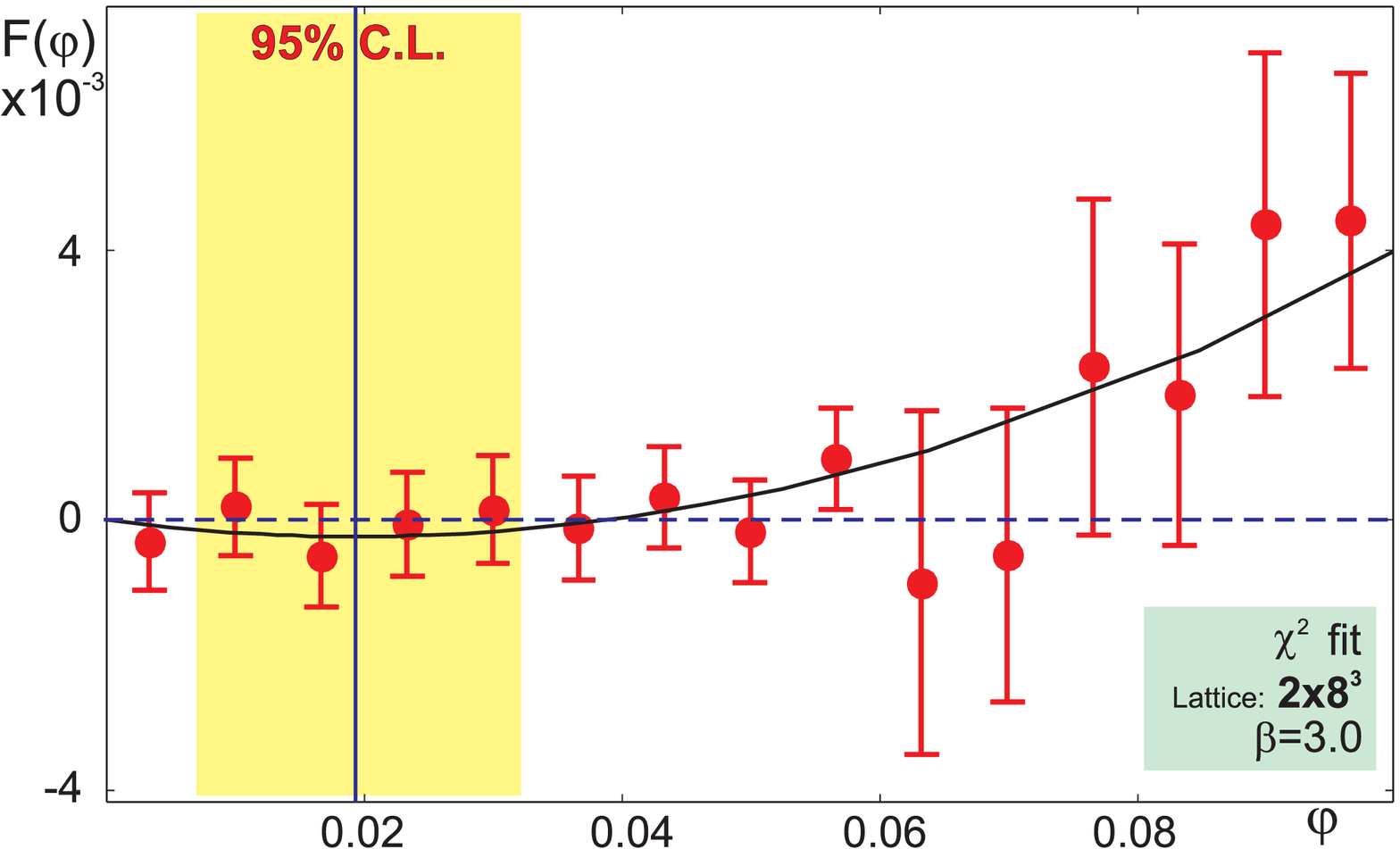}\hskip 1cm
\includegraphics[bb=0 0 581 352,width=0.45\textwidth]{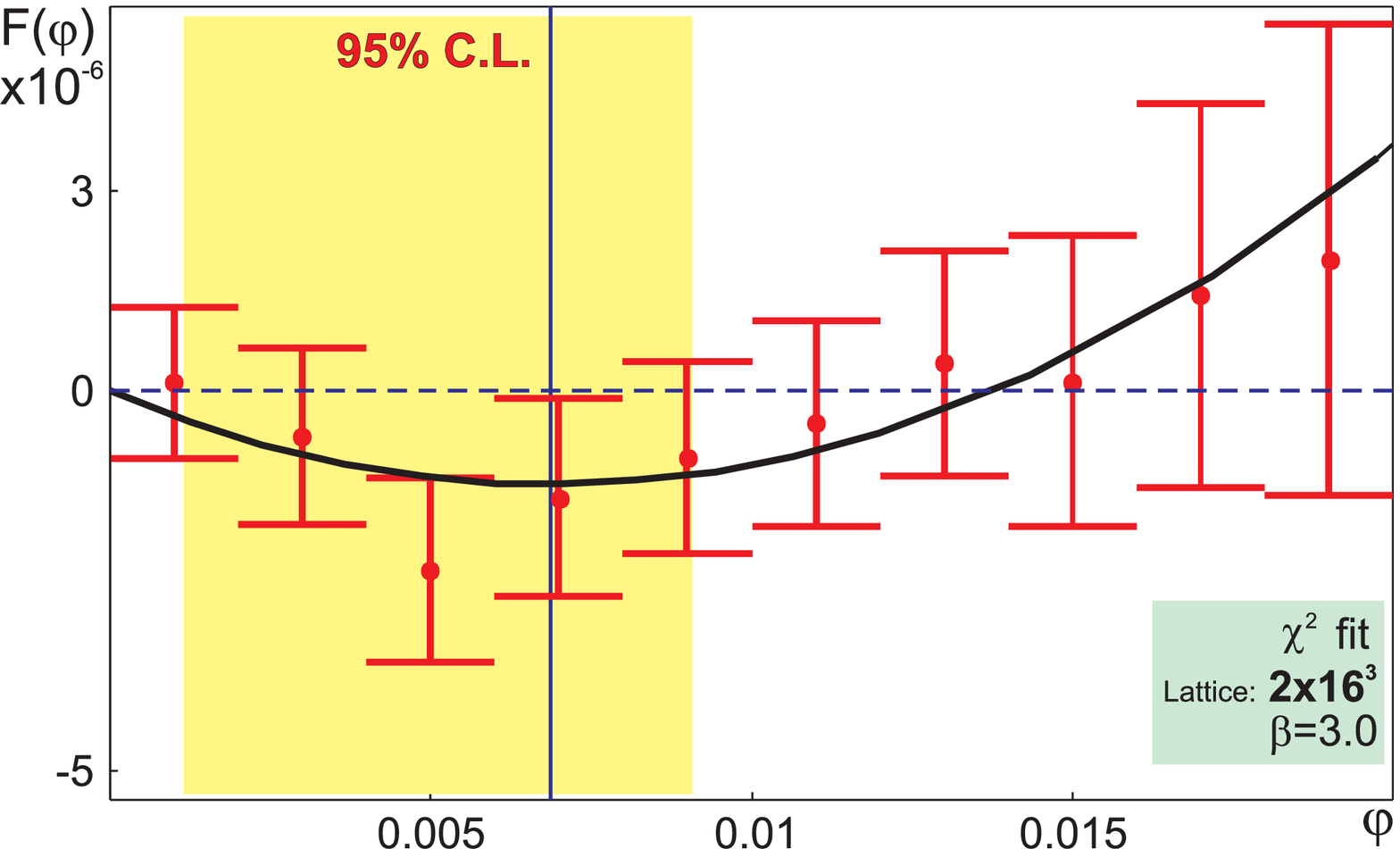}
\caption{$\chi^2$-fit of the free energy density on lattice $2\times 8^3$ (left) and $2\times 16^3$ (right) for {\small $\beta=3.0$} (yellow regions describe the {\small $\varphi_{min}=0.019^{+0.013}_{-0.012}$ and $\varphi_{min}=0.0069^{+0.0022}_{-0.0057}$}, at the 95\% confidence level).}
\label{figure:minfit}
\end{center}
\end{figure}

\begin{center}
\begin{tabular}{|c|c|c|c|} \hline \rule{0pt}{16pt}
 & $2\times 8^3$ & $2\times 16^3$ & $4\times 8^3$ \\\hline \rule{0pt}{16pt}
 $\beta=3.0$ & ${0.019^{+0.013}_{-0.012}}$ & $0.0069^{+0.0022}_{-0.0057}$ & $0.005^{+0.005}_{-0.003}$
 \\\hline \rule{0pt}{16pt}
 $\beta=5.0$ & ${0.020^{+0.011}_{-0.010}}$ & & \\\hline
\end{tabular}
\vskip 0.5cm ~\\\noindent\textsf{\bf{Table 1:}}
\mbox{\parbox[t]{0.84\hsize}{\textsf{The values of the generated fluxes $\varphi_{min}$ for different lattices (at the $95\%$ confidence level).}}}
\end{center}
The fit results are given in the Table~1. As one can see, $\varphi_{min}$ demonstrates the $2\sigma$-deviation from zero.
The 95\% C.L. area of the parameters $F_{min}$ ($b$ for the right figure) and $\varphi_{min}$ is represented in Fig.~\ref{figure:minfit}. The black cross marks the position of the maximum-likelihood values of $F_{min}$ ($b$ for the right figure) and $\varphi_{min}$. It can be seen that the flux is positively determined. The 95\% C.L. area becomes more symmetric with the center at the $F_{min}$, $b$ and $\varphi_{min}$ when the statistics is increasing. This also confirms the results of the fitting.

Thus, we see in this approach that spontaneous vacuum magnetization does take place at high temperature. This is in accordance with the results in Sect.3.

In Ref.~\cite{Demchik:2007ct} the spontaneous vacuum magnetization with accounting for the $A_0$-condensate at high temperature was investigated. This case has been studied in one-loop approximation in \cite{Starinets:1994vi, Meisinger:2002ji}. On the lattice, this parameter was introduced trough the Polyakov loop, as usually. The main result of these studies is that the spontaneous vacuum magnetization holds and the vacuum is stable for this case. Moreover, with taking into consideration of the condensate, the vacuum stability is increasing. Thus, at high temperature both condensates the constant $A_0$ and $B(T)$ are present.

\section{Magnetic mass on the lattice}
\qquad In this section we calculate the magnetic mass of the Abelian chromomagnetic field by using MC simulations. For that we, following Ref.~\cite{DeGrand:1981yx}, investigate the behavior of the average magnetic flux penetrating a lattice plaquette oriented perpendicular to the magnetic field direction. We introduce the classical magnetic field (\ref{potentialB}) on a lattice as before by applying the twisted boundary conditions. Note that in  Ref.~\cite{DeGrand:1981yx} the twist of the boundary conditions was applied to introduce the magnetic flux of the Dirac monopole.  Then, the magnetic mass of this non-Abelian magnetic field was measured by investigating the average plaquette values for the twisted and untwisted lattices. The main object of such type investigations is the difference (magnetic flux through a lattice plaquette perpendicular to the $OZ$ axis):
\begin{equation}\label{uflux}
 \langle U_{untwisted}\rangle - \langle U_{twisted}\rangle = f(m, L_s),
\end{equation}
which is fitted for each lattice geometry $L_t\times L_s^3$ by different functions $f(m, L_s)$.

Below we follow this general approach.  However, here we measure the magnetic mass of the Abelian field of interest, Eq.~(\ref{r22}).
The temperature is introduced as before through a lattice asymmetry in the temporal direction ($L_t<L_s$). The measurements were fulfilled for the value of $\beta = 2.6$ in the perturbation regime for the deconfinement phase. Lattices with $L_t=4$ and $L_s$ up to $32$ were used. To update the lattice, heat-bath algorithm with overrelaxation was used \cite{Creutz:1987xi}. To thermalize the system, up to $6000$ MC iterations were used. The plaquette average is calculated by averaging up to $10000$ working iterations.

To estimate the behavior of magnetic fields a large amount of simulation data must be prepared. Unfortunately, traditional computational resources are lack to perform the detail analysis. In our case, we use the General Purpose computation on Graphics Processing Units (GPGPU) technology allowing to study large lattices on personal computers. GPU programming model implemented here and some technical details on MC simulations on AMD/ATI graphics processing units (GPU) are given in Ref.~\cite{Demchik:2009ni}.

Distinguishing feature of the employed program model is that all necessary data for simulations are stored in GPU memory. GPU carries out intermediate actions and returns the results to the host program for final data handling and output. We avoid any data transfer during the run-time between the host program and kernels, to speed-up the execution process.

To generate the pseudo-random numbers for MC procedure, three different pseudo-random number generators are used: {\sf RANMAR}, {\sf RANLUX} and {\sf XOR128} \cite{Demchik:2010fd}. The last one allows to obtain the maximal performance but is not widely used in MC simulations. So, all the results were checked with the slower generators {\sf RANMAR} and {\sf RANLUX}.

The GPU-based MC program allows to calculate the difference (\ref{uflux}) for a wide interval of lattice geometries. Also, up to 1000 independent runs for each lattice size were performed in order to decrease the dispersion of the obtained values $f(m,L_s)$. The whole set of simulation data for different lattice geometries were fitted with the several functions which correspond to the different behavior of magnetic flux.

The results of fitting (fitting function, the values of the $\chi^2$-function corresponding to the $95\%$ confidence level and the obtained magnetic mass $m$) are shown in Table~2. The function $\frac{C}{r^2}$ corresponds to the magnetic flux tube formation ($r$ is the lattice size in the $X$ and $Y$ directions). The total magnetic flux through the lattice is conserved in this case. The function $\frac{C}{r^4}$ describes the Coulomb-like behavior and the function $\frac{C}{r^2} \exp(- m^2 r^2)$ is signaling the generation of the magnetic mass $m$ \cite{DeGrand:1981yx}. The functions $\frac{C}{r} \exp(- m r)$, $\frac{C}{r}$ can be related to the increase of the field strength with a temperature increase. This is
because the total magnetic flux through the lattice is growing faster than in the case of the magnetic flux tube formation.

As it follows from Table~2, the best fit function is $\frac{C}{r}\exp(-mr)$ with a small value of the magnetic mass $m=1.25\cdot 10^{-6}$. The value of $\chi^2$ function in this case is very close to the $m = 0$ situation and statistically these cases are indistinguishable. Really, the statistical errors are larger than the fitted value of $m$. Thus, from the carried out analysis we can conclude that the neutral component of the gluon field is not screened at high temperature like usual magnetic field. This result is in agreement with that of section 5 obtained in perturbation theory. Note that due to a large amount of data we have guarantied that the absolute value of errors is of $10^{2}-10^3$ times smaller than the value of the corresponding quantity.

\begin{center}
{\small \vskip 0.5cm
\begin{tabular}[c]{|l|c|c|c|}
 \cline{2-4} \multicolumn{1}{c}{~} & \multicolumn{3}{|c|}{\rule{0pt}{3ex}\bf Abelian field} \\
 \hline\rule{0pt}{3ex} {\bf Fit function} & $\chi^2$ & $C$ & $parameter$ \\\hline\hline
 \rule{0pt}{3ex}$C\exp(-mr)$ & $901.8$ & $0.063$ & $m = (2.44^{+0.06}_{-0.06})\cdot 10^{-2}$ \\
 \rule{0pt}{3ex}$C\exp(-m^2r^2)$ & $1924.4$ & $0.035$ & $m = (1.57^{+0.02}_{-0.02})\cdot 10^{-2}$ \\\hline
 \rule{0pt}{3ex}$C/r$ & $7.090$ & $0.911$ & \\
 \rule{0pt}{3ex}$C/r\exp(-mr)$ & $7.086$ & $0.912$ & $m = (1.25^{+52}_{-54})\cdot 10^{-6}$ \\
 \rule{0pt}{3ex}$C/r\exp(-m^2r^2)$ & $7.090$ & $0.911$ & $m^2 = (2.4^{+5951.2}_{-5784})\cdot 10^{-10}$ \\\hline
 \rule{0pt}{3ex}$C/{r^2}$ & $31400$ & $28.13$ & \\
 \rule{0pt}{3ex}$C/{r^2}\exp(-m^2r^2)$ & $7550$ & $18.26$ & $m^2 = -3.3\cdot 10^{-5} $\\\hline
 \rule{0pt}{3ex}$C/{r^4}$ & $159500$ & $248.9$ & \\
 \rule{0pt}{3ex}$C/{r^4}\exp(-m^4r^4)$ & $161000$ & $10.0$ & $m = 0.0$ \\\hline
\end{tabular}
\vskip 0.5cm ~\\\noindent\textsf{\bf{Table 2:}}
\mbox{\parbox[t]{0.84\hsize}{\textsf{Fit results for magnetic mass of Abelian magnetic field.}}} ~\\~ }
\end{center}

Interesting additional arguments in favor of spontaneous vacuum magnetization follow from the above measurements fulfilled. We observed that for the fitting function $f(m, L_s) = \frac{C}{r^2} $ corresponding to the magnetic flux tube formation the $\chi^2$ value is very large and entirely inconsistent with the data. But in the geometry of measurements it
describes the conservation of the magnetic flux introduced by the twist of the boundary conditions. The best fit functions $\frac{C}{r}, \frac{C}{r} \exp( - m r )$ with very small
(actually, zero) $m$ are signaling an increase of the mean magnetic field strength penetrating the plaquette perpendicular to the field direction. As a result, the flux through the whole $(X-Y)$ plain should increase. The only natural explanation is the spontaneous generation of the field inside the volume of the lattice.

\section{Magnetic field strength at $T_{ew}$}
In previous sections, we have shown that the spontaneous vacuum magnetization is realized at high temperature and determined the screening parameters $m_D(B, T)$ and $m_{magn.}$. Now, we are going to estimate in the SM the magnetic field strength at the temperatures close and higher the EWPT temperature, $T \geq T_{ew}$.

As it was shown in Sect.5, in the restored phase the hypermagnetic field $b_\mu = 0$ and the complete weak-isospin chromomagnetic field $A^{(3)}_\mu$ is generated. It is unscreened because its magnetic mass is zero. Thus, the field is a long-range one. It provides the coherence length $\lambda_B(T)$ to be sufficiently large. In fact, the field is a constant and occupies all a horizon scale at a given temperature, $\lambda_B(T) \sim R_{H(T)}$. Chromo(magnetic) fields of different types (color $SU(3)$, and others) can be spontaneously generated at high temperatures. This property could  be of  great importance for cosmology.

Now, we continue with describing a general field behavior related with the EWPT. In the restored phase, a scalar field condensate $\phi = 0$ and the constituent of the weak isospin field corresponding to the magnetic one is given by the expression
\begin{equation}\label{fieldT}
 B(T) = \sin \theta_w (T) B^{(3)}(T),
\end{equation}
where $B^{(3)}(T) = \tilde{B}$ is the strength of the field generated spontaneously. After the phase transition, the scalar condensate $\phi \not = 0$ and the field is partially screened.

To estimate the field strength $ \tilde{B}(T)$ in the restored phase at the EWPT temperature the total EP must be used. This can be best done numerically. To explain the procedure, we consider here the part of this potential accounting for the one-loop $W$-boson contributions. The high temperature expansion for the EP coming from charged vector fields is given in Eq.~(\ref{VW}). Assuming stability of the vacuum state, we calculate the value of chromomagnetic weak isospin field spontaneously generated at high temperature from Eqs.~(\ref{VW}) and (\ref{Vscalar}):
\begin{equation}\label{fieldT1}
 g \tilde{B}(T) = \frac{1}{16} \frac{g^4}{\pi^2} \frac{T^2}{(1 + \frac{5}{12} \frac{g^2}{ \pi^2} \log \frac{T}{\tau})^2 }.
\end{equation}
Here, $\tau$ is a reference temperature parameter.
This expression (and the complete one accounting for all the contributions) gives the field strength at any temperature $T \geq T_{ew}$. Such type formulas can be obtained for different models of particles.

Before educing a specific value of it, we describe how to relate this expression with the present day intergalactic magnetic field $B_0$. We first relate the expression (\ref{fieldT1}) with an magnetic field after symmetry breaking, and then take into account the scales of fields.
Let us introduce the standard parameters and definitions, $\frac{g^2}{4 \pi} = \alpha_s$, $\alpha = \alpha_s \sin \theta_w^2 $, $ \frac{(g')^2}{4 \pi} = \alpha_Y$ and $\tan^2 \theta_w(T) = \frac{ \alpha_Y(T)}{ \alpha_s(T)}$, where $\alpha$ is a fine structure constant. To find the temperature dependence of the Weinberg angle, the behavior of the hypercharge coupling $g'$ on the temperature has to be computed. From Eq.~(\ref{Vscalar}) it follows that this behavior is not trivial. The logarithmic temperature-dependent term is negative. But, as it is well known, in the asymptotically free models this sign must be changed to a positive value due to contributions of other fields. This particular value is model dependent. We will not calculate it in the present paper. Instead, for a rough estimate, we substitute the zero temperature number: $\sin^2 \theta_w(T) = \sin^2 \theta_w(0)= 0.23$.

For the given temperature of the EWPT, $T_{ew}$, the magnetic field is
\begin{equation}\label{B3T}
 B(T_{ew}) = B_0 \frac{T^2_{ew}}{T^2_0} = \sin \theta_w (T_{ew}) \tilde{B}(T_{ew}).
\end{equation}
This relation is the consequence of the assumption that for the field spontaneously generated at high temperature in the early Universe the magnetic flux conservation holds after the EWPT. It means that the field is "frozen" in plasma at large scales and the magnetic turbulence processes do not affect this behavior. Although this is most simple assumption, it requires a detailed discussion within the results obtained recently in magnetic hydrodynamics (see \cite{Banerjee:2004df, Kahniashvili:2009qi} and references therein).

Assuming $T_{ew} = 100 GeV = 10^{11} eV$ and $T_0 = 2.7 K = 2.3267 \cdot 10^{- 4} eV$, we obtain
\begin{equation}\label{Bew}
 B(T_{ew}) \sim 1.85\cdot ~10^{14} G.
\end{equation}
This value can be considered as an estimate of the magnetic field strength at the EWPT, if the standard model serves as a basic theory describing this phase transition. Hence, for the value of $X = \log \frac{T_{ew}}{\tau}$ we have the equation
\begin{equation}\label{Xew}
 B_0 = \frac{1}{2} \frac{\alpha^{3/2} }{\pi^{1/2} \sin^2 \theta_w } \frac{T^2_0}{(1 + \frac{5 \alpha}{3 \pi \sin^2 \theta_w } X)^{2}}.
\end{equation}
Since all the values are known, $\log \tau $ can be computed. After that the field strengths at different higher temperatures can be found. In fact, the main point in obtaining of these results is the assumption on the magnetic flux conservation and frozen in it in the plasma. An information on a particular model is implemented in the factor $\sin \theta_w (T_{ew})$ in Eq.~(\ref{B3T}).

Of course, our estimate is a rough one because of having ignored the temperature dependence of the Weinberg angle. To guess the value of the parameter $\tau$ we take the field strength $B_0 \sim 10^{- 9} G$, often used in the literature (see, for example, \cite{Pollock:2003iy}). In this case, from Eq.~(\ref{Xew}) we obtain $\tau \sim 300 eV$. For the present day value $B_0 \sim 10^{- 15}G$ this parameter is much smaller.

Let us compare now the value of the field strength (\ref{Bew}) with the one calculated directly from the EP for the SM in Ref.~\cite{Demchik:2001zq}. From Fig.~1 and Tab.~1 of that paper, we find
\begin{equation}\label{BewSM}
 B^{SM}(T_{ew}) \sim 10^{20} G,
\end{equation}
what is much larger than the value (\ref{Bew}) and just corresponds to the value of the present comoving field strength $B_0 \sim 10^{- 9} G$. This value was considered in numerous investigations as an upper bound on the intergalactic magnetic field strength in the Universe, before the recent discoveries \cite{Ando:2010rb, Neronov:1900zz, Essey:2010nd}. Note that the value of $B(T_{ew})$ calculated with expression (\ref{fieldT1}) is close to the estimate (\ref{BewSM}). Let us stress again that the field strength at higher temperatures will depend on the
particular model extending the standard one. Spontaneous vacuum magnetization in the minimal supersymmetric standard model has been investigated in Ref.~\cite{Demchik:2002ks}, and the field strength generated in this model is smaller as compared to the situation here considered. Also, Pollock \cite{Pollock:2003iy} has investigated this problem for the case of the Planck era, where magnetic fields of the order $B \sim 10^{52} G$ have been estimated.

An important consequence of Eq.~(\ref{fieldT1}) is that at high temperatures $T \ge T_{ew}$ the magnetic flux conservation does not hold. This follows from the logarithm term standing in the denominator. Therefore, at a given temperature $ T $ a specific magnetic field strength is generated due to vacuum polarization. This property is very important for what follows. It is a key point for the scenario which we are going to investigate. We assume for the following that intergalactic magnetic fields have been spontaneously generated in the hot Universe.

Note also that magnetohydrodynamical processes in the early Universe were investigated in numerous publications (see for references the review papers \cite{Widrow:2002ud, Kandus:2010nw}). Here, we mention only the important points for our consideration. The "frozen in" conditions can be realized for magnetic fields having the scales of largest turbulence eddies. After a free decay stage of the magnetized plasma evolution, the field can be considered as non-affected by turbulence \cite{Kahniashvili:2009qi}. In connection with these results, it is clear that the magnetic fields generated at the EWPT are not sufficient as such to produce a long-range correlated fields and some additional processes must be included. This is because even the field having the scale of the Hubble radius at temperature $T_{ew} \sim 100$ GeV is correlated at the comoving scale $l_0 \sim 10^{- 4}$ Mpc \cite{Kahniashvili:2009qi}. However, the scale of intergalactic magnetic fields was determined to be of the order $l_0^{ig} \sim 1$ Mpc \cite{Ando:2010rb, Neronov:1900zz, Essey:2010nd}. The fields generated at the EWPT can serve as seeds for long-range magnetic fields inside galaxies.

\section{Magnetic field scale}
Now we discuss the scale of the field generated in the restored phase \cite{Elizalde:2012kz}. This is a key point in relating expressions like Eqs.~(\ref{fieldT1}) or (\ref{BewSM}) with $B_0$. In our consideration, the ``frozen in" condition was used. So, let us discuss its applicability in more detail. Remind, if one assumes that after the EWPT the constant field $B(T_{ew})$ was frozen in the plasma at the Hubble scale, $R_H(T_{ew})$, then its comoving coherence scale at present has to be $\lambda_B(T_0) = 6 \cdot 10^{-4}$ pc \cite{Kahniashvili:2009qi}. This is much smaller than necessary.

We consider two, in fact related, ways to overcome such a difficulty \cite{Elizalde:2012kz}. The first is to take into consideration the reheating stage of the universe evolution. According to the concepts of modern cosmology \cite{Gorbunov:2011zzc}, this stage has existed just after inflation and was related with the latter causal stage. Just due to causality, the temperature in the Universe after this stage is the same, in all domains of space, which could even be uncorrelated in later moments of time.
Hence, at a given high temperature, $T$, a magnetic field generated due to vacuum polarization has the same strength $B(T)$ everywhere in the Universe. Formally, the field strength could have different directions, in either external or internal spaces. Different kind of (chromo)magnetic fields, of the type as in Eq.~(\ref{potentialB}), can be spontaneously generated. Their nature depends on a particular model considered and is therefore unknown, so far. But this is not essential for our consideration, here. The magnetic fields coherent on huge scales   have been present in the early Universe. The origin of this coherence is ensured by the properties of the solution to the field equations (Eq.~(\ref{potentialB})) and by causality at the inflationary epoch. The scales of the coherent field domains could be estimated on grounds of the gauge invariance. This idea, due to Feynman, was put in force in gluodynamics with the goal to determine possible magnetic vacuum structures \cite{Feynman:1981ss}. Namely, to find a gauge invariant vacuum, on the basis of gauge non-invariant solutions (such as Eq.~(\ref{potentialB})), one can consider a domain structure ensuring gauge invariance when a corresponding boundary is going around. At $T \not = 0$, this point requires further investigation.

Most of fields generated in the early Universe decouple and are screened at some energy (temperature) scales, when the corresponding scalar condensates have broken the background symmetries. So, the only unbroken symmetry at the EWPT remains the $SU(2)_{ew}\times U(1)$ one. After the EWPT, when spontaneous magnetization stops, this field cannot be included in turbulent processes generated by the transition. This is because the scale of the field, Eq.~(\ref{potentialB}), is already much larger than any largest eddy of turbulence. As it is usually believed, the size of a typical eddy is estimated as the inverse mass of the particles appearing after the transition \cite{Vachaspati:1991nm}. Thus, the field evolves in accordance with the metric expansion and is implemented in a hot plasma, thus fulfilling the magnetic flux conservation law. And it finally results in the present day intergalactic magnetic fields which could be correlated on $\sim 1$ Mpc scales. Note that an essential information on the processes that take place after the EWPT, obtained in the framework of magnetohydrodynamics, is given in Refs.~\cite{Kahniashvili:2009qi} and \cite{Subramanian:1997gi}.

Another possible scenario is based on the stochastic processes considered already by Hogan \cite{Hogan:1983zz} in connection with the magnetic fields generated at first-order EWPT. A possible mechanism of field generation in that case was proposed by Vachaspati \cite{Vachaspati:1991nm}. In the former paper, it was pointed out that magnetic fields correlated on large scales can be produced not only through causal processes but also by a stochastic random walk mechanism, if the magnetic lines generated in some domain of space ``forget" about their origin. The field strength developed on large scales by this process (due to ``straightening" of entangled magnetic fields) can be estimated as $B_N \sim B/\sqrt{N}$, where $N$ counts the number of domains, with the field $B$ of a given size, crossed by a magnetic line. The correlation length $\lambda_B$ in this case can be much larger than the $R_H(T)$. It can be estimated as $\lambda_B(T) \sim N R_H(T)$. In Ref.~\cite{Hogan:1983zz}, it has been also noticed that this mechanism is not applicable to the early Universe, the reason being because magnetic lines do not penetrate freely though the plasma. This is really the case, if the general properties of the plasma are taken into account. However, this is not the case if spontaneous vacuum magnetization is accounted for. In fact, at a given temperature, each uncorrelated domain of space having a Hubble radius $R_{H}(T)$ is filled up with a constant magnetic field $B(T)$, described by the potential (\ref{potentialB}). Its orientation in both external and internal spaces is arbitrary. Hence, a stochastic behavior of the field lines and the appearance of magnetic fields having large correlation lengths $\lambda_B(T) \ge R_H(T)$ are expected. After the EWPT, these fields evolve as in the previous case.

Note that, in both scenarios, all the fields generated at the inflation epoch \cite{Giovannini:2012tx} are washed out by the vacuum polarization and leave no remnants at present. The reheating stage becomes more important. In closing this section, we also remind that the long-range nature of the Abelian spontaneously generated magnetic fields is ensured by their zero magnetic mass, what renders these fields unscreened, as is the case for usual $U(1)$ magnetic fields. The difference, however, is essential because the former fields appear due to vacuum polarization and the latter ones need currents to be produced.

\section{Discussion and conclusions}
We here summarize our main results. We have determined the characteristic parameters of chromomagnetic fields at high temperature - Debye's and magnetic masses - in analytic and numeric approaches and obtained the coinciding results. Qualitatively, we showed  that the Debye mass of gauge fields in strong magnetic fields at high temperature is decreased as compared to the zero field case and therefore  electric fields are more long range ones. The magnetic mass of Abelian chromomagnetic fields was found to be  zero. So, these fields have to occupy  all the space as usual $U(1)$ magnetic field. This   property is very important for cosmology. In particular, for generation of long rang magnetic fields in the early Universe.

The spontaneous vacuum magnetization process eliminates the magnetic flux conservation principle at high temperature. Vacuum polarization is responsible for the value of the field strength $B(T)$ at each temperature and serves as its source. Physically, the magnetization is a consequence of the large magnetic moment of charged gauge bosons (gyromagnetic ratio $\gamma = 2$). This is one of the basic properties of matter. In fact, at high temperature, due to spin interaction of the virtual pairs, heat is converted into an ordered coherent vacuum state. This effect was not taken into consideration in previous studies of the early Universe.

We have shown here that, at finite temperature and after symmetry breaking, a scalar field condensate suppresses the magnetization. Hence, it follows that the actual nature of the model extending the standard one is {\it not} that essential at sufficiently low temperatures, when the decoupling of the heavy gauge fields has occurred. From this one can conclude, in particular, that the vacuum polarization ``washes out '' the relics of the magnetic fields generated at very high temperature or at the inflation stage. This is because different kinds of magnetic fields existed at high temperatures were screened with temperature lowering, and the magnetic flux conservation does not hold at $T \geq T_{ew}$. After the EWPT, when the solution $\phi \not = 0$ is realized, the spontaneous vacuum magnetization stops and the field evolves further according to the magnetic flux conservation law. The correlation length of the field is very large. At the EWPT, the field can be considered as constant magnetic field having the strength $B(T_{ew}) \sim 10^{14}$ G. This value is not large and the field not essentially influences various processes happening after the transition. This is because, according to the estimates in ~\cite{Vachaspati:1991nm}, the stochastic magnetic fields generated in a first order phase transition are much stronger, $B \sim 10^{20} - 10^{23}$ G. The scale of these fields is of the order of inverse particle mass, $ \lambda \sim 1/ g \phi$.

Recall that after reheating, the temperature in the hot Universe is the same everywhere (even in the domains of space which are uncorrelated in later periods of time). Analogous statement concerns the values of $B(T)$ which are strictly related with the temperature due to the vacuum polarization. Most magnetic fields existed at high temperatures were screened with the temperature lowering when the corresponding symmetries have been spontaneously broken. The only magnetic field at the temperatures close $T_{ew}$ is the $SU(2)_{ew}$ component described by the potential (\ref{potentialB}).

The present value of the intergalactic magnetic field strength, $B_0$, we related with the field in the restored phase. Because of the zero magnetic mass for Abelian magnetic fields, there is no problem for the generation of fields having large coherence scales. In our estimates, we have assumed that, basically, the field is of the order of the horizon scale, $\lambda_B(T) \sim R_{H(T)}$. This seems reasonable because, at a given temperature, the field $B(T) = const$, generated due to vacuum polarization, occupies all space. Then coherence on scales exceeding that of the horizon can be produced. Such large scale fields are not influenced by turbulent processes happening after the EWPT. They are frozen in the plasma and evolve according to the magnetic flux conservation law. In this scenario, a large scale domain structure is also permissible, what requires additional consideration.

As we have found above, the field strengths at the EWPT temperature, estimated with account to the present-day value of the intergalactic magnetic field strength, $B_0 \sim 10^{-15}$ G, Eq.~(\ref{Bew}), or either directly from the vacuum magnetization in the standard model, Eq.~(\ref{BewSM}), differ in six orders of magnitude. This huge deviation can be explained by the different scales of the fields considered. Let us check this possibility by using the second of the scenarios proposed in the previous section, for large scale field generation. We used the usual relation between the scale factor and the temperature,
\begin{equation}\label{a-T}
 \frac{a(T_{ew})}{a(T_{0})} = \frac{T_{0}}{T_{ew}},
\end{equation}
taken at the EWPT epoch, and the present-day parameters, $T_{ew}= 100 ~GeV = 10^{11} ~ eV$, $T_0 = 2.3267 \cdot 10^{- 4} ~eV$. If one now assumes that $ ~\lambda_B(T_{}) \sim a(T_{})$, then from (\ref{a-T}) it follows that $\lambda_B(T_0) = 6 \cdot 10^{-4}$ ps. On the other hand, if one takes $\lambda_B(T_0) = 1 $ Mpc, the value $\lambda_B(T_{ew}) = 2.33 \cdot 10^{- 15}$ Mpc is obtained. At the same time, the horizon size is $a(T_{ew})= 1.27 \cdot 10^{-24} $ Mpc, thus, $\lambda_B(T_{ew}) >> a(T_{ew})$. Now, following an idea of Hogan \cite{Hogan:1983zz}, we relate the size of the correlated field with the random walk process. At $T_{ew}$, we have $\lambda_B(T_{ew}) = N a(T_{ew})$, hence, we get roughly $\sqrt{N} = 3 \cdot 10^4$, and for the field strength ``straightened" on the $N$-domain scale, $B_N \sim \frac{B(T_{ew})}{\sqrt{N}}$ \cite{Hogan:1983zz}. Therefore, accounting for the field strength value calculated for the standard model, Eq.~(\ref{BewSM}), we obtain $B_{ls}(T_{ew}) \sim 3 \cdot 10^{15}$ G (the subscript in $B_{ls}$ means ``large scale"). This value is close the value $B_{ls}(T_{ew}) \sim 2 \cdot 10^{14}$ G estimated in Eq.~(\ref{Bew}). The remaining discrepancy can be explained in two ways. First, and obviously, as due to the roughness of our estimate. Second, and more radically, by the necessity of substituting the standard model with another one. The latter point will be discussed in more detail below.

Let us note a number of properties of the field under consideration, and compare them with the ones usually discussed in applications of magnetohydrodynamics to the early Universe. Here, we follow Refs.~\cite{Kahniashvili:2009qi} and \cite{Subramanian:1997gi} which are close to our analysis. First, we mention that the field energy density, $\rho_B = \frac{B^2}{2}$, is proportional to $g^6 T^4$, what is much smaller than the radiation energy density, $\sim T^4$. Thus, the BBN condition (see \cite{Kahniashvili:2009qi}), $\rho_B/\rho_{rad.} << 1$, is fulfilled. Second, as numerical simulations show \cite{Kahniashvili:2009qi}, the turbulent process in the early Universe with magnetic field included is slower than in the laboratory. Turbulence can include a large scale field at the level of the largest eddies. For large scale fields, the free decay stage is important. At this epoch, which is strongly dependent on the initial conditions \cite{Subramanian:1997gi}, turbulence is significantly decreasing, and after this very brief stage the field is not affected by turbulence any more. It is just frozen in the plasma. As we have shown, the spontaneous vacuum magnetization is stopped when the first-order phase transition ends. That is, just after the free decay stage. Fields of this type cannot be influenced by turbulence and thus it is reasonable to believe that after the EWPT the field evolves according to Eq.~(\ref{B3T}). Note also that these fields are non-helical ones.

Now, we compare our results with those of ~\cite{Pollock:2003iy}, where spontaneous vacuum magnetization at high temperature was applied to estimate the field strength at the Planck era. In that paper, to estimate the value $B(T)$ the heterotic superstring theory $E_8 \times E_8^{'}$ was considered as the basic model. At the Planck era, the magnetic field strength has been estimated to be of order $\sim 10^{52} G$. In contrast to our considerations, it was assumed there that the magnetic field approximately scales as $B \sim T^2$. That is, vacuum magnetization was taken into account only at the very first moments of the evolution. Further, recent results deriving a zero magnetic mass for the Abelian chromomagnetic fields also change the picture of the magnetized early Universe substantially. According to this, the created magnetic fields existed on the horizon scales. They were switched of at some mass scales. The processes of decoupling were also not taken into consideration in ~\cite{Pollock:2003iy}. Thus, it is impossible there to relate the magnetic field $B_0$ with the magnetic fields generated at high temperatures.

In the used approach, knowing the particular properties of the extended model, it is possible to estimate the field strengths at any temperature. This can be done for different schemes of spontaneous symmetry breaking (restoration) by taking into account the fact that, after the decoupling of some massive gauge fields, the corresponding magnetic fields are screened. Thus, the higher the temperature, the larger the number of strong long-range magnetic fields of different types that will be generated in the early Universe.
To estimate their field strengths at higher temperatures, one has to take into consideration a number of features proper to the standard model and its particular extension at play. First, we note that quarks possess both electric and color charges. Therefore, there is a mixing between the color and usual magnetic fields owing to the quark loops. Second, there are peculiarities related with the particular content of the extended models. For example, in the Two-Higgs-Doublet standard model the contribution $\sim (g B)^{3/2} T$ in Eq.~(\ref{VW}) is exactly canceled by the corresponding term in Eq.~(\ref{Vscalar}), because of the four charged scalar fields entering the model. They interact with gauge fields with the same coupling constant. However, in this model the doublets interact differently with fermions. This changes the effective couplings of the doublets with the gauge fields and results in non-complete cancelations. As a result, a suppression of the spontaneously created magnetic field is expected in this model. In principle, one should be able to explain, in this way, the discrepancy in the field strengths as discussed above. Note also that in the minimal supersymmetric standard model the field $B(T)$ generated at high temperature is smaller then in the standard model \cite{Demchik:2002ks}. So, formally, it can also be considered as a candidate for explanation of intergalactic magnetic fields. However, this case is not well studied, now. There can be other peculiarities which may influence the high temperature phase of the Universe.

As conclusion we note that the spontaneous vacuum magnetization at high temperature could find other interesting applications in cosmology. Another phenomenons of interest may be the deconfinement phase transition with accounting for both magnetic $B(T)$ and color chromomagnetic fields $B_c(T)$.  The situation here may differ from described above because gluons are massless and spontaneously generated color magnetic field has to exist till deconfinement temperature $T_d$. These require further investigations.

\section*{Acknowledgements}
Authors indebted Michael Bordag, in collaboration with whom numerous results of the present paper have been obtained, for useful discussions and suggestions. They also thankful Emilio Elizalde for reading the manuscript and advices. Authors are grateful Alexey Gulov for useful discussions and suggestions.

\end{document}